\documentclass[useAMS,usenatbib]{mn2e}
\usepackage{ifthen}
\usepackage{graphicx}
\usepackage{float}
\usepackage{amssymb}
\usepackage{wasysym}
\usepackage{pifont}
\usepackage[usenames]{color}

\def\jref@jnl#1{{\rm#1}}

\def\aj{\jref@jnl{AJ}}                   
\def\araa{\jref@jnl{ARA\&A}}             
\def\apj{\jref@jnl{ApJ}}                 
\def\apjl{\jref@jnl{ApJ}}                
\def\apjs{\jref@jnl{ApJS}}               
\def\ao{\jref@jnl{Appl.~Opt.}}           
\def\apss{\jref@jnl{Ap\&SS}}             
\def\aap{\jref@jnl{A\&A}}                
\def\aapr{\jref@jnl{A\&A~Rev.}}          
\def\aaps{\jref@jnl{A\&AS}}              
\def\azh{\jref@jnl{AZh}}                 
\def\baas{\jref@jnl{BAAS}}               
\def\jrasc{\jref@jnl{JRASC}}             
\def\memras{\jref@jnl{MmRAS}}            
\def\mnras{\jref@jnl{MNRAS}}             
\def\pra{\jref@jnl{Phys.~Rev.~A}}        
\def\prb{\jref@jnl{Phys.~Rev.~B}}        
\def\prc{\jref@jnl{Phys.~Rev.~C}}        
\def\prd{\jref@jnl{Phys.~Rev.~D}}        
\def\pre{\jref@jnl{Phys.~Rev.~E}}        
\def\prl{\jref@jnl{Phys.~Rev.~Lett.}}    
\def\pasa{\jref@jnl{PASA}}               
\def\pasp{\jref@jnl{PASP}}               
\def\pasj{\jref@jnl{PASJ}}               
\def\qjras{\jref@jnl{QJRAS}}             
\def\skytel{\jref@jnl{S\&T}}             
\def\solphys{\jref@jnl{Sol.~Phys.}}      
\def\sovast{\jref@jnl{Soviet~Ast.}}      
\def\ssr{\jref@jnl{Space~Sci.~Rev.}}     
\def\zap{\jref@jnl{ZAp}}                 
\def\nat{\jref@jnl{Nature}}              
\def\iaucirc{\jref@jnl{IAU~Circ.}}       
\def\aplett{\jref@jnl{Astrophys.~Lett.}} 
\def\apspr{\jref@jnl{Astrophys.~Space~Phys.~Res.}}
\def\bain{\jref@jnl{Bull.~Astron.~Inst.~Netherlands}} 
\def\fcp{\jref@jnl{Fund.~Cosmic~Phys.}}  
\def\gca{\jref@jnl{Geochim.~Cosmochim.~Acta}}   
\def\grl{\jref@jnl{Geophys.~Res.~Lett.}} 
\def\jcap{\jref@jnl{JCAP}}      %
\def\jcp{\jref@jnl{J.~Chem.~Phys.}}      
\def\jgr{\jref@jnl{J.~Geophys.~Res.}}    
\def\jqsrt{\jref@jnl{J.~Quant.~Spec.~Radiat.~Transf.}}
\def\memsai{\jref@jnl{Mem.~Soc.~Astron.~Italiana}}
\def\nphysa{\jref@jnl{Nucl.~Phys.~A}}   
\def\physrep{\jref@jnl{Phys.~Rep.}}   
\def\physscr{\jref@jnl{Phys.~Scr}}   
\def\planss{\jref@jnl{Planet.~Space~Sci.}}   
\def\procspie{\jref@jnl{Proc.~SPIE}}   

\def\r0{r_{_0}}
\def\cm2og{\,\mathrm{cm}^2\,\mathrm{g}^{-1}}
\def\omegaR{\omega_{_\mathrm{R}}}
\def\omegaI{\omega_{_\mathrm{I}}}

\title[galaxy halo stability]{
Galaxy stability within a self-interacting dark matter halo
}
\author[C. J. Saxton]{Curtis J. Saxton$^{1}$\thanks{E-mail:
cjs2@mssl.ucl.ac.uk (CJS);
}, 
\\
$^{1}$Mullard Space Science Laboratory, University College London,
Holmbury St Mary, Surrey RH5 6NT, UK
}

\begin{document}

\date{Accepted 2012 December 20.  Received 2012 December 11; in original form 2012 September 24}

\pagerange{\pageref{firstpage}--\pageref{lastpage}} \pubyear{2011}

\maketitle

\label{firstpage}

\begin{abstract}
This paper investigates spheroidal galaxies comprising
   a self-interacting dark matter halo (SIDM)
   plus de~Vaucouleurs stellar distribution.
These are coupled only
   via their shared gravitational field,
   which is computed consistently from the density profiles.
Assuming conservation of mass, momentum and angular momentum,
   perturbation analyses reveal
   the galaxy's response to radial disturbance.
The modes depend on
   fundamental dark matter properties,
   the stellar mass,
   and the halo's mass and radius.
The coupling of stars and dark matter stabilises some haloes
   that would be unstable as one-fluid models.
However the centrally densest haloes are unstable,
   causing radial flows of SIDM and stars
   (sometimes in opposite directions).
Depending on the dark microphysics,
   some highly diffuse haloes are also unstable.
Unstable galaxies might shed their outskirts or collapse.
Observed elliptical galaxies appear to exist in the safe domain.
Halo pulsations are possible.
The innermost node of SIDM waves
   may occur within ten half-light radii.
Induced stellar ripples may also occur at detectable radii
   if higher overtones are excited.
If any SIDM exists,
   observational {\em skotoseismology} of galaxies
   could probe DM physics, measure the sizes of specific systems,
   and perhaps help explain peculiar objects
   (e.g. some shell galaxies, and the growth of red nuggets).
\end{abstract}

\begin{keywords}
	dark matter
	---
	galaxies: elliptical and lenticular, cD
	---
	galaxies: kinematics and dynamics
	---
	galaxies: structure
\end{keywords}

\section{Introduction}

For decades it has been observed that visible matter in galaxies and clusters
   is insufficient to bind them gravitationally
	\citep[e.g.][]{zwicky1937,freeman1970,roberts1973}.
The extra gravity of invisible ``dark matter'' (DM) is invoked,
   but its properties are uncertain,
   with many particle physics conjectures offering candidates.
On Earth,
   DM particle detection experiments
   differ in their findings:
   some report annually modulated signals
   \citep[e.g.][]{bernabei2008,aalseth2011a,aalseth2011b,angloher2011};
   others see nothing above their expected background
   \citep[e.g.][]{angle2008,aprile2010,akerib2006,ahmed2009}.
Reconciling them may require extra physics:
   inelastic or composite forms of dark matter,
   or extra internal degrees of freedom
   \citep[e.g.][]{smith2001,chang2009,alves2010,kaplan2010a,kaplan2010b}

Large-scale cosmic filaments and voids emerge
   in $N$-body computer simulations using gravity alone
   \citep[e.g.][]{frenk1983,melott1983,springel2006}
{\ifthenelse{\isundefined{\rainbow}}{}{\color{MidnightBlue}}%
   but there is observational tension
   concerning smaller, self-bound systems.
The predicted abundances of dwarf galaxies
   exceed those seen
   in many types of environments
\citep[e.g.][]{klypin1999,moore1999a,donghia2004,tikhonov2008,
	carlberg2008,zavala2009,peebles2010}.
The biggest simulated subhaloes appear too massive
   to match real satellite galaxies
   \citep{boylan2011,boylan2012}.
}%
Collisionless DM haloes develop
   singular central density {\em cusps}
   \citep[e.g.][]{gurevich1988,dubinski1991,nfw1996}.
On the contrary,
   observations of diverse galaxy classes indicate or allow
   a dark density profile
   that is smooth and nearly uniform within a kpc-scale inner {\em core},
   \citep[e.g.][]{flores1994,moore1994,burkert1995,salucci2000,kelson2002,
	kleyna2003,goerdt2006,
        gentile2004,deblok2005,thomas2005,kuzio2006,
        gilmore2007,weijmans2008,oh2008,nagino2009,inoue2009,
        deblok2010,pu2010,murphy2011,memola2011,walker2011,
	jardel2012,agnello2012,salucci2012,lora2012}.
Galaxy cluster haloes should be centrally compressed
   by the steep density profile of cooling gas,
   but nontheless many examples are definitely or possibly cored
   \citep[e.g.][]{sand2002,sand2008,ettori2002,halkola2006,halkola2008,
        voigt2006,rzepecki2007,zitrin2009,newman2009,newman2011,richtler2011}.

To remedy (or defer) these DM crises,
{\ifthenelse{\isundefined{\rainbow}}{}{\color{MidnightBlue}}%
   many
   }%
   astrophysicists invoke energetic ``feedback''
   by starbursts or active galaxies:
   explosions and outflows violent enough to flatten the dark cusps
   and denude small subhaloes of baryons
   \citep[e.g.][]{navarro1996,gnedin2002,mashchenko2006,
		peirani2008a,governato2010}.
These events would have happened early enough to be difficult to observe today.
The details would involve radiative shocks
   in magnetised multiphase interstellar media
   that are thermally and dynamically unstable.
The relevant scales are numerically unresolvable in present simulations.
Results depend on phenomenological recipes
   for sub-grid processes,
   and the implementations are more varied and elaborate than any DM physics.
For reasons of resolution alone,
   the proposition that feedback cures the DM crises
   cannot be decisively evaluated for at least another decade.
{\ifthenelse{\isundefined{\rainbow}}{}{\color{MidnightBlue}}%
Regardless, feedback may be
   energetically incapable of removing cusps
   whilst leaving realistic stellar populations in dwarf galaxies
   \citep{penarrubia2012}.
}

An alternative approach extends
   standard particle physics phenomenologies:
   taking the dark matter crises as signatures of
   dynamically important physics in the dark sector.
In ``SIDM'' theories,
   dark matter is self-interacting
   yet electromagnetically neutral.
Pressure support or thermal conduction
   could prevent cusps,
   while drag and ablation reduce subhaloes
   \citep[e.g.][]{flores1994,moore1994,spergel2000,firmani2000,
	vogelsberger2012,rocha2012}.
{\ifthenelse{\isundefined{\rainbow}}{}{\color{MidnightBlue}}%
SIDM interactions entail direct inter-particle scattering
   (with specific crossection $\varsigma$)
   possibly enhanced by short-range Yukawa-type forces,
   or a long-range ``dark electromagnetism''
   \citep[e.g.][]{ackerman2009,loeb2011}.
Collisionless plasma supported by ``dark magnetism''
   is conceivable.
}%
Alternatively, the SIDM might consist of boson condensates
   of a scalar field
   \citep[e.g.][]{peebles2000}.
Plausible early-universe evolution
   and cosmic structures
   emerge in cosmogonies with fluid-like SIDM
   \citep[e.g.][]{moore2000,stiele2010,stiele2011}
   or scalar field dark matter
   \citep[e.g.][]{matos2001,suarez2011}.
There were early fears that SIDM haloes
   would form {\em sharper} cusps via gravothermal collapse
   \citep[e.g.][]{burkert2000,kochanek2000}.
Later analyses put this beyond a Hubble time,
   and found two safe regimes
   that suppress thermal conductivity:
   weaker scattering (rarely per orbit);
   and stronger (fluid-like) cases with tiny mean free paths
   \citep{balberg2002b,ahn2005,koda2011}.
{\ifthenelse{\isundefined{\rainbow}}{}{\color{MidnightBlue}}%
We will refer to the
   $\varsigma\la1\cm2og$ and $\varsigma\ga200\cm2og$
regimes as ``weak-SIDM'' and ``strong-SIDM'' respectively.
}

The ubiquity of dark cores supports SIDM,
   but there have been circumstantial counter-arguments.
It was argued
   that observed halo ellipticities imply low collisionality
   \citep[e.g][]{miralda2002,buote2002}.
Such limits cannot apply to systems that are elliptical due to
   tidal disturbance, rotation or non-radial pulsations.
Newer SIDM simulations reject the spherical prediction
   and imply a looser limit $\varsigma<1\cm2og$
   \citep{peter2012}.
Other papers inferred tight limits on SIDM collisionality,
   from the growth of supermassive black holes
   \citep[$\varsigma<0.02\cm2og$:][%
	]{ostriker2000,hennawi2002}
   and the survival of cluster members
   \citep[$<0.35$ or $\ga10^4\cm2og$,][]{gnedin2001}.
However those investigations were in a way self-fulfilling,
   because they assumed {\em cuspy} initial conditions,
   with a strong thermal inversion at the centre.
This choice was inconsistent with cored profiles expected from SIDM.

Gravitational lensing models of merging galaxy clusters
   give conflicting signs about DM physics:
   sometimes the inferred DM seems collisionless,
   following galaxies and separating from gas
   \citep[$\varsigma<1.25\cm2og$, e.g.][]{clowe2006,randall2008}.
In other systems DM seems collisional, following the gas
   or separating from galaxies
   \citep[e.g.][%
   ]{mahdavi2007,jee2012,williams2011}.
The derived limits depend on
   assumed projected geometries and histories of unique systems;
   these are each open to specific counter-explanations and rebuttals.
For instance if the acclaimed ``Bullet Cluster''
   were truly a direct-hit merger
   then ram pressure and shocks would quench star formation.
This is not observed \citep{chung2009}.

\cite{saxton2008}
   modelled galaxy clusters
   in which the halo could possess extra heat capacity
   (e.g. composite SIDM particles).
Stationarity in the presence of gas inflows
   requires a central accretor exceeding some minimum mass.
For the range of DM properties giving plausible
   cluster and core profiles,
   a part of the inner halo 
   teeters towards local gravitational instability,
   implying likely collapse to form supermassive black holes of realistic sizes.
\cite{saxton2010} modelled elliptical galaxies
   comprising a polytropic halo plus coterminous collisionless stars.
Observed stellar kinematics of elliptical galaxies
   were fitted by the same class of DM 
   that gives the most realistic clusters.

This paper is a further theoretical study of spheroidal galaxies.
Their stability properties
   will imply constraints on the end-points of galaxy evolution.
If galaxies do possess SIDM haloes
   then their oscillatory modes will
   open the future possibility of {\em skotoseismology}
   --- acoustic diagnosis of dark matter haloes that pulsate.
The paper's organisation is as follows:
   \S\ref{s.formulation}
	formulates the stationary model and perturbations;
   \S\ref{s.results}
	describes the oscillatory solutions and instabilities;
   \S\ref{s.discussion}
	discusses applications and interpretations;
   \S\ref{s.conclusions}
	concludes.

\section{Formulation}
\label{s.formulation}

This paper investigates
   spherical galaxies comprising
   self-interacting dark matter and collisionless stars
   in near-equilibrium configurations.
The SIDM halo is polytropic, adiabatic, non-singular and radially finite
   (\S\ref{s.halo}).
The system is isolated: unconfined and unexcited by any external medium.
Stellar orbits are isotropically distributed,
   in an observationally motivated density profile
   (\S\ref{s.stars}).
Oscillations and instabilities are calculated
   in response to radial perturbations
   (\S\ref{s.perturbations}--\S\ref{s.stellar.perturbations}).
Mass displacements are non-uniform,
   and at any location the two constituents may be displaced
   with differing amplitudes and phases.
The perturbed variables are non-singular at the origin,
   and the outer boundary of the halo is free
   (\S\ref{s.bc}).

\subsection{dark matter halo profile}
\label{s.halo}

For each mode in which ideal collisional gas particles
   are free to move,
   the energy per particle is
   ${\frac12}kT={\frac12}\mu\sigma_{_\mathrm{d}}^2$
   where $k$ is Boltzmann's constant,
   $T$ is temperature,
   $\mu$ is the particle mass
   and $\sigma_{_\mathrm{d}}$
   is a characteristic (one-dimensional) velocity dispersion.
If SIDM gas has $F$ effective thermal degrees of freedom
   that are locally in equipartition
   (due to collisionality or collective effects),
   then
   the specific heat capacities at constant volume
   and pressure are
   $c_\mathrm{V}={\frac12}Fk$
   and $c_\mathrm{P}=c_\mathrm{V}+k$.
The ratio of specific heats
\begin{equation}
	\gamma\equiv{{c_{_\mathrm{P}}}\over{c_{_\mathrm{V}}}}=1+{2\over{F}}
	\ .
\end{equation}
From the first law of thermodynamics,
   adiabatic processes
   imply a polytropic equation of state
   for the dark pressure:
\begin{equation}
	P_{_\mathrm{d}} \equiv \rho_{_\mathrm{d}}\, \sigma_{_\mathrm{d}}^2
	= s\, \rho_{_\mathrm{d}}^\gamma
	\ ,
\label{eq.state}
\end{equation}
   or (equivalently) the mass density is
	$\rho_{_\mathrm{d}}=Q\, {\sigma_{_\mathrm{d}}}^F$
   where $s$ is a pseudo-entropy,
   and $Q=s^{-F/2}$ is a generalised phase-space density.
In adiabatic processes, $s$ and $Q$ remain constant.

The micro-physical meanings of ``$F$'' deserve further remarks.
Point-like particles moving in three spatial dimensions
   have $F=3$, due to translational kinetic energy.
Composite particle substructure raises $F$
   (e.g. rotational and vibrational modes).
A diatomic gas has $F=5$ because two parts
   are moving in 3D space, at a fixed separation.
Incompressible fluids are more constrained and $F=0$.
For a relativistic or radiation dominated plasma, $F=6$.
If particles were massive enough (short wavelength)
   to traverse hidden spatial dimensions,
   then extra translational modes would raise $F$.
An isothermal medium
   has $F\rightarrow\infty$.

Polytropic fluids also emerge
   in some non-classical theories of dark matter.
Scalar field and boson condensate models usually imply $F=2$,
   but possibly other $F$ values
   depending on an index in the interaction potential
   \citep[e.g.][]{goodman2000,peebles2000,arbey2003,
		boehmer2007,harko2011a,harko2011c,chavanis2011,robles2012}.
This paper is agnostic about which particle theory
   causes the value of $F$,
   which should ultimately be constrained by observations.

The radial structure of a spherical halo
   is determined by its momentum equation,
\begin{equation}
	\rho_{_{\rm d}}\,{{\,\mathrm{d}^2r}\over{\,\mathrm{d}t^2}}
	+
	{{\mathrm{d}P_{_{\rm d}}}\over{\mathrm{d}r}}
	= -{{G\left({ m_\bigstar+m_{_{\rm d}} }\right)\rho_{_{\rm d}}
	} \over{r^2}} \ . 
\label{eq.momentum}
\end{equation}
The stellar mass enclosed at radius $r$ is
   $m_\bigstar=m_\bigstar(<r)$
   (see \S\ref{s.stars}).
The corresponding enclosed dark mass is obtained
   by numerical integration of 
\begin{equation}
	{{\mathrm{d} m_{_{\rm d}}}\over{\mathrm{d}r}}
	= 4\pi r^2 \rho_{_{\rm d}}
	\ .
\label{eq.dark.mass0}
\end{equation} 
In hydrostatic equilibrium,
   the condition (\ref{eq.momentum}) becomes
\begin{equation}
	{{\mathrm{d}\sigma_{_{\rm d}}^2}\over{\mathrm{d}r}}
	= -{2\over{F+2}}
		{{G\left({m_\bigstar+m_{_{\rm d}} }\right) } \over{r^2}}
	\ .
\label{eq.ode.sigma}
\end{equation}
Neglecting any compact central mass (e.g. black hole),
   a non-singular boundary condition applies at the origin
   ($\mathrm{d}\sigma_{_{\rm d}}^2/\mathrm{d}r|_{r=0}=0$).
The equilibrium dark halo is adiabatic,
   and its pseudo-entropy $s$ is spatially uniform.
This may occur in nature if a galaxy halo is well mixed,
   or if $s$ is set by universal constants of the SIDM particle.

\begin{figure*}
\begin{center}
\includegraphics[width=164mm]{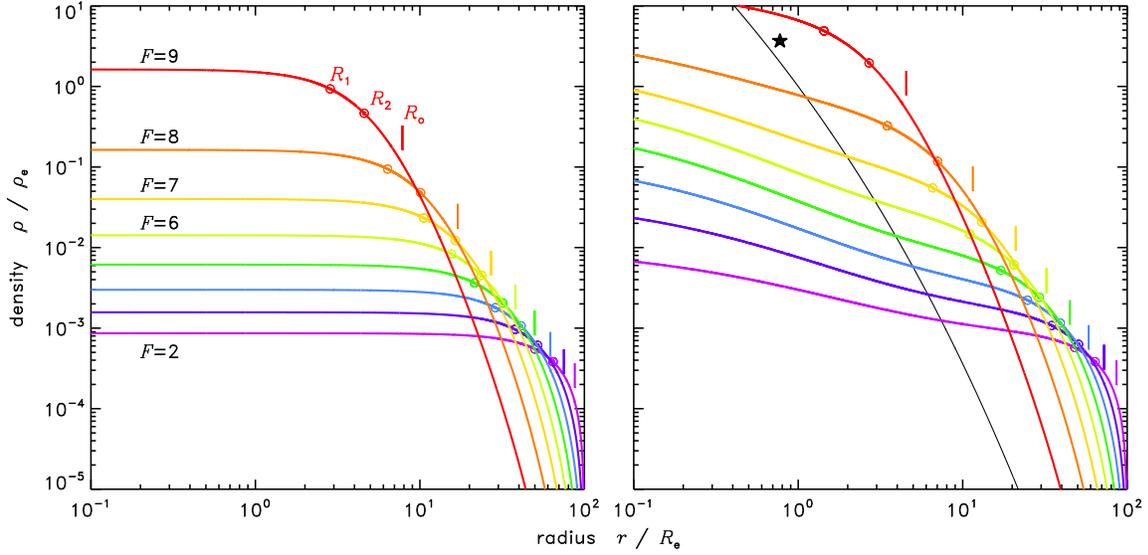}
\end{center}
\caption{%
Radial density profiles of
   equal-mass haloes with $F=2,\ldots,9$ (coloured)
   truncating at the same radius ($R=100R_\mathrm{e}$).
The left panel shows pure polytropes;
   the right panel includes a stellar profile
   (\S\ref{s.stars}, ``$\bigstar$'')
   massing $1/24$ times the halo.
The halo density index is $-1$ and $-2$ at the $\Circle$ and $\odot$ symbols
   (radii $R_1$ and $R_2$)
   just outside the uniform core.
Ticks locate the peak circular orbit velocity ($R_\mathrm{o}$).
}
\label{fig.cores}
\end{figure*}

For $-2<F<10$ a halo of finite mass
   truncates ($\rho_{_\mathrm{d}}\rightarrow0$)
   at a finite outer radius $R$.
For other $F$,
   any finite mass equilibrium model has infinite radius:
   it boils off into the void,
   or fails to condense from the cosmic background in the first place.
The high limit ($F=10$) is a generalised \cite{plummer1911} model,
   with infinite radius and finite mass.
{\ifthenelse{\isundefined{\rainbow}}{}{\color{MidnightBlue}}%
Haloes may also seem radially infinite in some SIDM simulations with $F=3$
   \citep[e.g.][]{rocha2012}.
This may be due to ongoing supersonic infall of neighbouring objects,
   or because the scattering is weak enough
   to leave the outskirts collisionless.
If the SIDM ``strength'' is increased
   (greater scattering $\varsigma$,
   or long-range ``dark magnetism'' effects)
   then polytropic behaviour affects more of the halo.  
Beyond some threshold strength, a definite surface must occur.
\cite{rocha2012} note a temperature drop at large radii
   in their largest $\varsigma$ simulation.
This is attributable to incipient truncation.

The value of $F$ determines the natural ratio of core to system radii:
   for greater $F$ the core is smaller
   (e.g. compare cases in Figure~\ref{fig.cores}).
If dark mass dominates,
   the circular orbit velocity profile
   ($V=\sqrt{Gm_{_\mathrm{d}}/r}$)
   peaks at a radius $R_\mathrm{o}$
   which occurs at
   $R_\mathrm{o}/R\approx0.750, 0.0782$
   when $F=3$ and $F=9$ respectively.
In these cases, the dark density index drops to
   $\mathrm{d}\ln\rho_{_\mathrm{d}}/\mathrm{d}\ln{r}=-1$
   at radii
   $R_1/R\approx0.379, 0.0285$;
   and index $-2$ at
   $R_2/R\approx0.520, 0.0459$.
In other words, a strong-SIDM core with $F=9$
   is radially smaller than a tenth of a $F=3$ core.
Appendix~\ref{appendix.radii}
   lists standard radii for other $F$ cases.
Adding stars distorts these ratios
   but the general trend with $F$ persists.
Figure~\ref{fig.cores} (right panel)
   shows how a stellar component
   can shrink and steepen the dark core.
For comparison,
   the ``scale radius'' of simulated CDM haloes \citep{nfw1996}
   is $R_2$,
   and $R_\mathrm{o}\approx2.163R_2$.
For the \cite{burkert1995}
   cored profile with scale radius $r_\mathrm{b}$,
   the key radii are
   $R_1\approx0.657r_\mathrm{b}$,
   $R_2\approx1.521r_\mathrm{b}$,
   and $R_\mathrm{o}\approx3.244r_\mathrm{b}$.
Simulated SIDM cores fitted by \cite{rocha2012}
   ($r_\mathrm{b}\approx0.7R_2$)
   appear consistent with Burkert's parameterisation.
The ratios $R_1:R_2:R_\mathrm{o}$ are similar
   for Burkert models and $F=3$ polytropes,
   but less similar for other $F$.

These ideal core sizes are upper limits.
In collisional SIDM, weaker $\varsigma$
   delays the formation of full-sized cores.
Current SIDM simulations focus on the $F=3$ case,
   which predicts cores larger than those observed,
   unless the scattering crossection is kept small
   \citep[$\varsigma<1\cm2og$, e.g.][]{yoshida2000b,
		arabadjis2002,katgert2004,vogelsberger2012,rocha2012}.
The alternative is to vary $F$:
   the range of $7\la F<10$ results in galaxy clusters
   with realistic $\sim10^1$--$10^2$kpc cores
   \citep{saxton2008}.
In principle, haloes with $F\ga7$ could produce cores of realistic radius
   for arbitrarily large $\varsigma$.
Such $F$ values also provide $R_\mathrm{o}/R$
   small enough to be consistent with observations
   of large flat regions in disc galaxies' rotation curves.
The choice of $F\ga7$ is therefore an astrophysically preferable subdomain,
   but for the sake of generality this paper
   considers a wider interval, $2\le F\le9$.
}%

\subsection{stellar mass profile}
\label{s.stars}

The assumed stationary distribution of the stars
   resembles the empirical \cite{sersic1968} form,
   where the surface density at projected radius $X$ is
\begin{equation}
	\Sigma_\bigstar(X) \propto
	\exp\left[{-b(X/R_\mathrm{e})^{1/n}}\right]
	\ ,
\label{eq.sersic}
\end{equation}
   and $n$ is the ``S\'{e}rsic index.''
The ``effective radius'' $R_\mathrm{e}$ encircles half the starlight.
Assuming that the stellar mass-to-light ratio is constant,
   there exist exact deprojections of (\ref{eq.sersic})
   that express the 3D density profile
	\citep[e.g.][]{mazure2002},
   but these involve exotic special functions.
We will instead use the approximate deprojection of
   \cite{prugniel1997}, with density and mass profiles
\begin{equation}
	\rho_\bigstar = \rho_{\rm e}\ 
	\exp\left[{
		b-b\left({r/R_\mathrm{e}}\right)^{1/n}
	}\right]\ (r/R_\mathrm{e})^{-p}
	\ \ \ \ \mbox{and}
\label{eq.ps.density}
\end{equation}
\begin{equation}
	m_\bigstar(<r) = 4\pi n b^{n(p-3)} \mathrm{e}^b\ 
	\rho_{\rm e}R_{\rm e}^3\ 
	\Gamma\left[{
		n(3-p), b\left({r/R_{\rm e}}\right)^{1/n}
	}\right]
	\ ,
\label{eq.star.mass0}
\end{equation}
where $\Gamma(a,u)$ is the lower incomplete gamma function. 
The index $n$ determines $b$ and $p$
   \citep{limaneto1999,cotti1999,marquez2000}.
Adopting $n=4$ to match the classic \cite{devaucouleurs1948} profile,
   the values are $b\approx7.669$ and $p\approx0.8510$.
For this standard choice,
   the total stellar mass is
   $M_\bigstar=45.79\rho_\mathrm{e}R_\mathrm{e}^3$,
   and that within the half-light radius is
   $m_\bigstar(1R_\mathrm{e})=19.03\rho_\mathrm{e}R_\mathrm{e}^3$.

The frequency scale of the stars will be abbreviated as
\begin{equation}
	\omega_\mathrm{e}\equiv\sqrt{G\,\rho_\mathrm{e}}
	\ .
\end{equation}
Given the properties of large elliptical galaxy like M87,
   $\rho_\mathrm{e}\sim0.04\,m_\odot\,\mathrm{pc}^{-3}$ and
   $\omega_\mathrm{e}\approx0.01~\mathrm{Myr}^{-1}$
   \citep{murphy2011}.
For a compact elliptical such as M32,
   $\rho_\mathrm{e}\sim14\,m_\odot\,\mathrm{pc}^{-3}$ and
   $\omega_\mathrm{e}\sim0.3~\mathrm{Myr}^{-1}$
   \citep{mateo1998,choi2002}.
For ``red nuggets'' at high redshift,
   $\rho_\mathrm{e}\ga3\,m_\odot\,\mathrm{pc}^{-3}$ and
   $\omega_\mathrm{e}\la0.05~\mathrm{Myr}^{-1}$
   \citep[e.g.][]{toft2012,vandesande2011}.
If stars were the only component present,
   the keplerian period at $r=1R_\mathrm{e}$
   would be $t_\mathrm{k}=1.440/\omega_\mathrm{e}$,
   and the angular frequency $\kappa_\mathrm{e}=4.363\omega_\mathrm{e}$.

It is convenient to write the density gradient index as
\begin{equation}
	\alpha_\bigstar
	\equiv-{{\mathrm{d}\ln\rho_{_\bigstar}}\over{\mathrm{d}\ln r~}}
	= 
			p + {{b\,(r/R_\mathrm{e})^{1/n}}\over{n}}
		\ ,
\label{eq.star.alpha}
\end{equation}
   noting that $\alpha_{_\bigstar}\rightarrow p$ near the origin.
In central regions a modified radial coordinate,
   $z\equiv r^{1-p}$,
   is useful to soften terms involving the density,
\begin{equation}
	{{\mathrm{d}}\over{\mathrm{d}z}}\ldots
	={{r^p}\over{1-p}}
	{{\mathrm{d}}\over{\mathrm{d}r}}\ldots
	\ ;
\label{eq.ddz}
\end{equation}
   as the $r^p$ factor cancels the singular part
   of $\rho_{_\bigstar}$ in any ODEs.
%

The stellar velocity dispersions may in principle differ
   in the three orthogonal directions:
   radial ($\sigma_r$),
   polar ($\sigma_\theta$)
   and azimuthal ($\sigma_\phi$).
We will consider spherical models
   in which the transverse velocity dispersions are equal,
   $\sigma_\theta=\sigma_\phi\equiv\sigma_\perp$.
The radial and transverse velocity dispersions are related
   by defining an anisotropy profile,
\begin{equation}
	\beta=\beta(r)\equiv
	1-{{\sigma_\perp^2}\over{\sigma_r^2}}
	\ ,
\end{equation}
   \citep[e.g.][]{osipkov1979,merritt1985,mamon2005}.
The radial ``pressure'' of stars is
\begin{equation}
	P_\bigstar \equiv \rho_\bigstar\sigma_r^2
	\ ,
\label{eq.star.pressure}
\end{equation}
   and its profile
   is calculated from a spherical Jeans equation for momentum,
   expressed in a frame
   co-moving with the mean radial velocity ($v_\bigstar$)
   of a mass ``packet'' of stars
   \citep{jeans1915,binney1987},
\begin{equation}
	{{\mathrm{d}P_\bigstar}\over{\mathrm{d}r}}
	=
	g \rho_\bigstar
	-{{\rho_\bigstar}\over{r}}\left({
		2\sigma_r^2-\sigma_\theta^2-\sigma_\phi^2
	}\right)
	-
	\rho_\bigstar{{\mathrm{d}v_\bigstar}\over{\mathrm{d}t}}
	\ ,
\label{eq.jeans.general}
\end{equation}
{\ifthenelse{\isundefined{\rainbow}}{}{\color{MidnightBlue}}%
where $g\equiv-G(m_\bigstar+m_\mathrm{d})/r^2$
is the gravitational field.
}

\subsection{acoustic \& orbital timescales}
\label{s.acoustic}

The acoustic crossing time of the dark matter
   from the origin to the outer surface
   is obtained by radial integration of
\begin{equation}
	{{\mathrm{d}t_{\rm d}}\over{\mathrm{d}r}}
	={{1}\over\sqrt{\gamma\sigma_{\rm d}^2}}
	\ .
	\label{eq.sonic.dark}
\end{equation}
For the stars, the analogous ``acoustic'' time is:
\begin{equation}
	{{\mathrm{d}t_\bigstar}\over{\mathrm{d}r}}
	\equiv
	\sqrt{ {\partial\rho_\bigstar}\over{\partial P_\bigstar} }
	=
	\sqrt{
	\left.{{\mathrm{d}\rho_\bigstar}\over{\mathrm{d}r}}
	\middle/
	{{\mathrm{d}P_\bigstar}\over{\mathrm{d}r}}\right.
	}
\ .
\label{eq.sonic.star}
\end{equation}
Assuming stationarity,
   then substituting (\ref{eq.star.alpha}) and (\ref{eq.jeans.general})
   simplifies (\ref{eq.sonic.star}):
   e.g. if $\beta=0$ then
   $\mathrm{d}t_\bigstar/\mathrm{d}r=\sqrt{\alpha_\bigstar/g_0r_0}$.
From dimensional considerations,
   the crossing time should scale with the halo radius
   as $t_{\rm d}\sim R^{3/2}$.
Appendix~\ref{s.stationary}
   shows examples of the exact relations.
When calculating the natural modes of the galaxy,
   we should expect oscillations to scale
   in proportion to the angular frequencies 
   $\Omega_\mathrm{d}=2\pi/t_\mathrm{d}$
   and $\Omega_\bigstar=2\pi/t_{_\bigstar}$.
Once the eigenfrequencies
{\ifthenelse{\isundefined{\rainbow}}{}{\color{MidnightBlue}}%
   ($\omega$)}
   of a galaxy are known,
   one can compare them to local keplerian frequencies,
\begin{equation}
	\kappa = \sqrt{{G(m_\bigstar+m_{\rm d})}\over{r^3}}
	\ .
\end{equation}
In regions where $\kappa\gg\omega$,
   perturbations should only affect stellar orbits adiabatically.
Wherever $\kappa\ll\omega$, tidal shocking may occur
   when disturbances exceed the linear regime.

\subsection{perturbative displacements and gravitation}
\label{s.perturbations}

For each of the variables describing local properties
   of the dark matter or stars,
we define a first order perturbation.
For instance for a density variable we have
\begin{equation}
	\rho = \rho_0 + \widetilde{\rho}\,\varepsilon e^{i\omega t}
		  = \rho_0\left({
	  		1 + \lambda_\rho\,\varepsilon e^{i\omega t}
	  }\right)
\end{equation}
   where the subscript ``0'' denotes the stationary solution;
   the ``$\sim$'' part describes the response to perturbation;
   the $\lambda$ function describes the perturbation in relative terms
   (e.g. $\lambda_\rho\equiv \widetilde{\rho}/\rho_0$).
{\ifthenelse{\isundefined{\rainbow}}{}{\color{MidnightBlue}}%
Perturbed variables such as $\lambda_\rho$ and $\widetilde{\rho}$
   are complex.
}
The small constant $\varepsilon$ is the scale of perturbation;
Higher order terms are discarded ($\varepsilon^2\rightarrow0$).
The frequency of perturbation is complex
   ($\omega=\omega_{_\mathrm{R}}+i\omega_{_\mathrm{I}}$)
   with the real part describing oscillations
   and the imaginary part describing the tendency to growth or decay.
   
As in \cite{eddington1918} we shall consider lagrangian perturbations.
For a mass shell that sits at radius $\r0$ in equilibrium,
   we write its displacement and velocity under perturbation:
\begin{equation}
   r = \r0 + \widetilde{r}\,\varepsilon e^{i\omega t}
   \ ,
\end{equation}
\begin{equation}
   v = i\omega\widetilde{r}\,\varepsilon e^{i\omega t}
   \ .
\end{equation}
The dark matter and stars at any $\r0$
   may be displaced differently,
   so we treat them in two slightly different lagrangian frames,
   and distinguish their coordinates carefully:
\begin{eqnarray}
r_{\mathrm{d}}
&\hspace{-3mm}=\hspace{-3mm}&
\r0+\widetilde{r}_\mathrm{d}\,\varepsilon e^{i\omega t}
=\r0\left({1+\lambda_{r_\mathrm{d}}\,\varepsilon e^{i\omega t}}\right)
\\
r_\bigstar
&\hspace{-3mm}=\hspace{-3mm}&
\r0+\widetilde{r}_\bigstar\,\varepsilon e^{i\omega t}
=\r0\left({1+\lambda_{r_\bigstar}\,\varepsilon e^{i\omega t}}\right)
\\
v_\mathrm{d}
&\hspace{-3mm}=\hspace{-3mm}&
i\omega\widetilde{r}_\mathrm{d}\,\varepsilon e^{i\omega t}
=i\omega \r0\lambda_{r_\mathrm{d}}\,\varepsilon e^{i\omega t}
\\
v_\bigstar
&\hspace{-3mm}=\hspace{-3mm}&
i\omega\widetilde{r}_\bigstar\,\varepsilon e^{i\omega t}
=i\omega \r0\lambda_{r_\bigstar}\,\varepsilon e^{i\omega t}
\end{eqnarray}
{\ifthenelse{\isundefined{\rainbow}}{}{\color{MidnightBlue}}%
As the perturbed variables are complex,
   the mass constituents may be displaced with different phases
   (i.e. it is possible that
   $\arg({\widetilde{r}_\mathrm{d}})\neq\arg({\widetilde{r}_\bigstar})$).
}
Time derivatives are straightforward,
   e.g. for density,
\begin{equation}
	{{\mathrm{d}\rho}\over{\mathrm{d}t}}
{\ifthenelse{\isundefined{\rainbow}}{}{\color{MidnightBlue}}%
	={{\mathrm{d}}\over{\mathrm{d}t}}
	\widetilde{\rho}\varepsilon e^{i\omega t}
	}
	=i\omega\rho_0\varepsilon e^{i\omega t}
	\ .
\end{equation}
Expressions for the perturbation of each radial gradient term
   depend on whether we consider the comoving frame
   of the stars or the dark matter.
Let's abbreviate ``$+$'' and ``$-$''
   for ``$\bigstar$'' and ``$\mathrm{d}$'' subscripts.
Mass conservation requires that
$\rho_\pm r_\pm^2\mathrm{d}r_\pm = \rho_{\pm_0} \r0^2\mathrm{d}\r0$,
and then in the $\pm$ frame we have the operators
\begin{equation}
	{{\mathrm{d}}\over{\mathrm{d}r_\pm}}
	=
	{{\mathrm{d}}\over{\mathrm{d}\r0}}
	+\left({
		\lambda_{\rho_\pm}+2\lambda_{r_\pm}
	}\right)\varepsilon e^{i\omega t}
	{{\mathrm{d}}\over{\mathrm{d}\r0}}
	\ .
	\label{eq.gradient}
\end{equation}
The gravitational field provides the only coupling between components.
Consistent evaluation of its perturbations (in both frames)
   requires special care.
The stellar and dark shells that rest at $\r0$ in the stationary solution
   will feel the same field strength there ($g_0$)
   but the perturbed versions of $g$ differ
   since the shells oscillate through displacements
   with different phase and amplitude.
Define
\begin{equation}
	g_\pm = g_0 + \widetilde{g}_\pm\,\varepsilon e^{i\omega t}
	\ .
\label{eq.def.g}
\end{equation}
For newtonian gravity, we have $g_0=-Gm_0/\r0^2$
{\ifthenelse{\isundefined{\rainbow}}{}{\color{MidnightBlue}}%
   with $m_0=m_0(<\!\r0)=m_{0_-}+m_{0_+}$
}
   being the total mass contained within $\r0$ in equilibrium.
Poisson's equation applies at all orders of perturbation,
   which implies
\begin{equation}
	{{\mathrm{d}g_\pm}\over{\mathrm{d}r_\pm}}
	=
	-{{2g_\pm}\over{r_\pm}}
	-4\pi G\left({
		\rho_+ + \rho_-
	}\right)
\ .
\end{equation}
Combining (\ref{eq.gradient}) and (\ref{eq.def.g})
   gives the perturbed part of $\widetilde{g}_\pm$
   as felt by the $\pm$ component.
Denoting
$\widetilde{g}_\pm\equiv-G\widetilde{m}_\pm/\r0^2$
   gives
\begin{eqnarray}
	G{{\mathrm{d}\widetilde{m}_\pm}\over{\mathrm{d}\r0}}
	\hspace{-3mm}&=\hspace{-3mm}&
	-2g_0\r0\left({
		\lambda_{\rho_\pm} +3\lambda_{r_\pm}
	}\right)
\nonumber\\&&
	-4\pi G \r0^2\left[{
		2\rho_{0_\pm}\lambda_{r_\pm}
		+\rho_{0_\mp}\left({
			\lambda_{\rho_\pm} -\lambda_{\rho_\mp}
			+2\lambda_{r_\pm}
		}\right)
	}\right]
\ ,
\nonumber\\
\label{eq.mass.perturbed}
\end{eqnarray}
which has the advantage of being nonsingular 
and
{\ifthenelse{\isundefined{\rainbow}}{}{\color{MidnightBlue}}%
$\widetilde{m}_+=\widetilde{m}_-=0$
}
at the origin.

\subsection{dark matter perturbations}

The dark matter perturbations are considered adiabatic
   ($s$ is spatially and temporally constant).
The equation of state links the pressure and density perturbations:
\begin{equation}
	\lambda_{P_\mathrm{d}}
	\equiv
	{{\widetilde{P}_\mathrm{d}
	}\over{P_{0_\mathrm{d}}}}
	=
	\gamma\,\lambda_{\rho_\mathrm{d}}
	\equiv
	\gamma\,
	{{\widetilde{\rho}_\mathrm{d}
	}\over{\rho_{0_\mathrm{d}}}}
	\ .
\end{equation}
Mass conservation then yields
\begin{eqnarray}
	{{\mathrm{d}\widetilde{r}_{_\mathrm{d}}}\over{\mathrm{d}\r0}}
	&\hspace{-3mm}=&\hspace{-3mm}
	-\lambda_{\rho_\mathrm{d}} -2\lambda_{r_\mathrm{d}}
	=
	-{{\widetilde{P}_{_\mathrm{d}}}\over{\gamma P_{0_\mathrm{d}}}}
	-2{{\widetilde{r}_{_\mathrm{d}}}\over{\r0}}
	\ ,
\label{eq.dark.mass.1}
\\
	{{\mathrm{d}\lambda_{r_\mathrm{d}}}\over{\mathrm{d}\r0}}
	&\hspace{-3mm}=&\hspace{-3mm}
	-{1\over{\r0}}\left({
		{{\lambda_{P_\mathrm{d}}}\over\gamma}
		+3\lambda_{r_\mathrm{d}}
	}\right)
	\ .
\label{eq.dark.mass.2}
\end{eqnarray}
The lagrangian equation for momentum (\ref{eq.momentum})
   decomposes under (\ref{eq.gradient})
   to give the hydrostatic condition,
\begin{equation}
	{{\mathrm{d}P_{0_\mathrm{d}}}\over{\mathrm{d}\r0}}
	=\rho_{0_\mathrm{d}}\,g_0
	\ ,
\end{equation}
   and a perturbation equation, writable in two forms:
\begin{eqnarray}
{{\mathrm{d}\widetilde{P}_{_\mathrm{d}}}\over{\mathrm{d}\r0}}
&\hspace{-3mm}=&\hspace{-3mm}
\rho_{0_{\rm d}}\left({
	\omega^2\r0\lambda_{r_\mathrm{d}}
	-2g_0\lambda_{r_\mathrm{d}}
{\ifthenelse{\isundefined{\rainbow}}{}{\color{MidnightBlue}}%
	+\widetilde{g}_\mathrm{d}
	}
	}\right)
	=\rho_{0_{\rm d}}g_0\,\Lambda_{P_{\rm d}}
	\ ,
\label{eq.dark.momentum.1}
\\
	{{\mathrm{d}\lambda_{P_\mathrm{d}}}\over{\mathrm{d}\r0}}
	&\hspace{-3mm}=&\hspace{-3mm}
	{{g_0}\over{\sigma^2_{0_{\rm d}}}}
	(\Lambda_{P_{\rm d}}-\lambda_{P_{\rm d}})
	\ ,
\label{eq.dark.momentum.2}
\label{eq.dark.momentum}
\end{eqnarray}
with the abbreviation
\begin{eqnarray}
	\Lambda_{P_\mathrm{d}}
	\equiv
	{{\omega^2\r0\lambda_{r_\mathrm{d}}
	-2g_0\lambda_{r_\mathrm{d}}
{\ifthenelse{\isundefined{\rainbow}}{}{\color{MidnightBlue}}%
	+\widetilde{g}_{_\mathrm{d}}
	}
	}\over{g_0}}
	={{\widetilde{m}_{_\mathrm{d}} }\over{m_0}}
	-\left({
		2+{{\omega^2}\over{\kappa^2}}
	}\right)\lambda_{r_{\rm d}}
	\ .
\label{eq.Lambda}
\end{eqnarray}
The presence of the stars is felt through the gravitational perturbation
{\ifthenelse{\isundefined{\rainbow}}{}{\color{MidnightBlue}}%
   $\widetilde{g}_{_\mathrm{d}}$
   }
   obtained from (\ref{eq.mass.perturbed}).

\subsection{stellar perturbations}
\label{s.stellar.perturbations}

For the collisionless stellar component,
   the perturbed mass conservation equation
   resembles the dark matter version,
   but without an equation of state to link density and pressure.
\begin{eqnarray}
	{{\mathrm{d}\widetilde{r}_{_\bigstar}}\over{\mathrm{d}\r0}}
	&\hspace{-3mm}=&\hspace{-3mm}
	-{{\widetilde{\rho}_{_\bigstar}}\over{\rho_{0_\bigstar}}}
	-2{{\widetilde{r}_{_\bigstar}}\over{\r0}}
	= -\lambda_{\rho_\bigstar} -2\lambda_{r_\bigstar}
	\ ;
\label{eq.star.displacement}
\\
	{{\mathrm{d}\lambda_{r_\bigstar}}\over{\mathrm{d}\r0}}
	&\hspace{-3mm}=&\hspace{-3mm}
	-{1\over{\r0}}\left({
		{\lambda_{\rho_\bigstar}}
		+3\lambda_{r_\bigstar}
	}\right)
	\ .
\label{eq.star.mass}
\end{eqnarray}

Next, we define perturbed forms of 
   the radial and transverse velocity dispersions of the stars,
\begin{equation}
	\sigma_r^2 = \sigma_0^2 
	+ \widetilde{\sigma^2}\,\varepsilon e^{i\omega t}
\end{equation}
\begin{equation}
	\sigma_\perp^2 = (1-\beta)\sigma_0^2 
	+ \widetilde{a^2}\,\varepsilon e^{i\omega t}
	\ ,
\end{equation}
{\ifthenelse{\isundefined{\rainbow}}{}{\color{MidnightBlue}}%
   where $\widetilde{\sigma^2}$ and $\widetilde{a^2}$ are the perturbations of
   the radial and transverse velocity dispersions respectively.
}
The stationary models
   in this paper assume isotropy ($\beta=0$ everywhere)
   although orbital anisotropies
   occur transiently during perturbations.
Nonetheless let's retain $\beta$ terms in the general formulation.
If the perturbations are radial then there is no torque.
If the specific angular momentum of a mass element is conserved
   during a radial displacement then each local value of
   $r^2\sigma_\perp^2$ stays constant,
   eliminating a variable:
\begin{equation}
	\widetilde{a^2} 
	= -{{2(1-\beta)\widetilde{r}_{_\bigstar}}\over{\r0}} \sigma_0^2
	= -2(1-\beta)\sigma_0^2\lambda_{r_\bigstar}
	\ .
\label{eq.torque}
\end{equation}
Expressing the radial pressure perturbation as
   $\widetilde{P}=\widetilde{\rho}\sigma_0^2+\rho_0\widetilde{\sigma^2}$,
   provides a convenient relation,
\begin{equation}
\widetilde{\sigma^2} = \sigma_{0}^2
	\left({
		\lambda_{P_\bigstar} - \lambda_{\rho_\bigstar}
	}\right)
	\ .
\end{equation}

The Jeans equation (\ref{eq.jeans.general})
   separates into stationary and perturbed parts,
\begin{equation}
	{{\mathrm{d}P_{0_\bigstar}}\over{\mathrm{d}\r0}}
	=
	g_0\rho_{0_\bigstar}
	-{{2\beta P_{0_\bigstar}}\over{\r0}}
	\ ,
\label{eq.jeans.stationary}
\end{equation}
\begin{eqnarray}
	{{\mathrm{d}\lambda_{P_\bigstar}}\over{\mathrm{d}\r0}}
	&\hspace{-3mm}=&\hspace{-3mm}
	{{
	\left({
		\omega^2\widetilde{r}_{_\bigstar}
{\ifthenelse{\isundefined{\rainbow}}{}{\color{MidnightBlue}}%
		+\widetilde{g}_\bigstar
		}
		-2g_0\lambda_{r_\bigstar}
		-g_0\lambda_{P_\bigstar}
	}\right)
	}\over{\sigma_{0_\bigstar}^2}}
	\nonumber\\&&
	-{{2}\over{\r0}}\left[{
		(2-5\beta)\lambda_{r_\bigstar}
		+(1-\beta)\lambda_{P_\bigstar}
		-\lambda_{\rho_\bigstar}	
	}\right]
	\ .
\label{eq.jeans.lambda}
\end{eqnarray}
Given another Jeans equation for evolution
   of the radial stellar pressure
   (Appendix~\ref{appendix.energy}),
\begin{eqnarray}
0&\hspace{-3mm}=&\hspace{-3mm}
	{{\mathrm{d}P_{_\bigstar}}\over{\mathrm{d}t}}
	+2v_{_\bigstar}^3{{\partial\rho_{_\bigstar}}\over{\partial r}}
	+\left({
		3P_{_\bigstar}+2\rho_{_\bigstar} v_{_\bigstar}^2
	}\right){{\partial v_{_\bigstar}}\over{\partial r}}
	\nonumber\\
	&&+{{2\rho_{_\bigstar} v_{_\bigstar}}\over{r}}\left({
		v_{_\bigstar}^2+2\sigma_r^2-\sigma_\perp^2		
	}\right)
	\ ,
\end{eqnarray}
{\ifthenelse{\isundefined{\rainbow}}{}{\color{MidnightBlue}}%
   and performing the perturbation expansion
   yields an equation in the stationary variables
   (which reduces to a tautology, $0=0$)
   plus an equation in perturbed variables.
}
The latter, first order equation is:
\begin{eqnarray}
0&\hspace{-3mm}=&\hspace{-3mm}
	i\omega\widetilde{P}_{_\bigstar}
	+3i\omega P_{0_\bigstar}
	{{\mathrm{d}\widetilde{r}_{_\bigstar}}\over{\mathrm{d}\r0}}
	+2i\omega
	(1+\beta)
	P_{0_\bigstar}\lambda_{r_\bigstar}
	\ .
\label{eq.jeans.mess}
\end{eqnarray}
Substituting (\ref{eq.star.displacement})
   into (\ref{eq.jeans.mess})
   leads to an algebraic relation,
\begin{equation}
	0 = \lambda_{P_\bigstar} -3\lambda_{\rho_\bigstar}
	+(2\beta-4)\lambda_{r_\bigstar}
\label{eq.star.closure}
\end{equation}
which eliminates the $\lambda_{\rho_\bigstar}$ terms in
(\ref{eq.star.displacement}),
(\ref{eq.star.mass})
and (\ref{eq.jeans.lambda}).

\subsection{boundary conditions}
\label{s.bc}

At the origin, by symmetry, there cannot be
   any oscillatory displacement of mass,
   and therefore
\begin{equation}
   \lim_{\r0\rightarrow0} \widetilde{r}_{_\mathrm{d}}
   =
   \lim_{\r0\rightarrow0} \widetilde{r}_{_\bigstar} =0
   \ .
\label{eq.ibc.r}
\end{equation}
In {\em relative} terms the offset is free;
   any small nonzero value of $\lambda_{r_\mathrm{d}}$ is acceptable
   at the inner boundary.
Regularity of the mass equations
   (\ref{eq.dark.mass.2})
   and (\ref{eq.star.mass})
   then gives:
\begin{equation}
	\lim_{\r0\rightarrow0}\lambda_{\rho_\bigstar}
	= -3\lambda_{r_\bigstar}
	\ ;
\label{eq.ibc.star1}
\end{equation}
\begin{equation}
	\lim_{\r0\rightarrow0}\lambda_{P_\mathrm{d}}
	= -3\gamma\lambda_{r_\mathrm{d}}
	\ .
\end{equation}
Combining (\ref{eq.star.closure}) and (\ref{eq.ibc.star1}) implies
\begin{equation}
	\lim_{\r0\rightarrow0}\lambda_{P_\bigstar}
	= -(5+2\beta)\lambda_{r_\bigstar}
	\ ,
\label{eq.ibc.star2}
\end{equation}
and $\widetilde{P}_\bigstar = P_{0_\bigstar}\lambda_{P_\bigstar}$
using the central pressure from the stationary model.
Analogously, the central pressure oscillation of dark matter is
   $\widetilde{P}_\mathrm{d}=P_{0_\mathrm{d}}\lambda_{P_\mathrm{d}}$.
Assuming that the stars and dark matter are comoving
   in the dense inner regions
   requires $\lambda_{r_\bigstar}=\lambda_{r_{\mathrm d}}$ at the origin.
This completes the set of inner boundary conditions.

As $\sigma_{0_\mathrm{d}}\rightarrow0$
   at the halo surface,
   (\ref{eq.dark.momentum.2}) usually tends towards singularity.
To avoid this,
   the natural modes must satisfy an outer boundary condition:
\begin{eqnarray}
	\lim_{r\rightarrow R}\left({
		\lambda_{P_\mathrm{d}}
	-\Lambda_{P_\mathrm{d}}
	}\right)
	=0
	\ .
\label{eq.obc}
\end{eqnarray}
Condition (\ref{eq.obc}) is only satisfied at $\omega$ eigenfrequencies.
The stellar mass profile lacks any finite outer boundary.
Therefore there are no outer boundary conditions applicable to
   the $\bigstar$ perturbed variables.

\subsection{numerical integration and solution}
\label{s.numerical}

{\ifthenelse{\isundefined{\rainbow}}{}{\color{MidnightBlue}}%
To calculate the stationary profiles,
   trial values are set for the pseudo-entropy constant $s$ 
   and the central velocity dispersion of dark matter
   ($T_0\equiv\sigma_{_{\rm d}}^2|_{r=0}$).
The system of coupled differential equations
   (\ref{eq.dark.mass0})--(\ref{eq.ode.sigma})
   is integrated numerically as an intial value problem in radius
   (using the {\sc rk4imp} and {\sc rk8pd} integrators from the
   {\em Gnu Scientific Library}\footnote{%
	{\tt http://www.gnu.org/software/gsl/}}).
In practice it is necessary to change the independent variable
   (the integration coordinate) partway through the integral.
Near the origin a softened radial coordinate
   (\ref{eq.ddz})
   improves accuracy in the stellar cusp.
In the outskirts it is more practical to use
   $\sigma_\mathrm{d}^2$ or
   $\ln(\sigma_{\rm d}^2)$
   as the independent variable,
   and integrate to a standard limit at
   $\sigma_{\rm d}^2=10^{-8}$ or $10^{-12}$.
(The chosen limit doesn't change any results,
   as long as it is consistent across all runs.)
At the outer boundary (radius $R$)
   the total dark mass is recorded
   ($M_{_{\rm d}}\equiv m_{_{\rm d}}|_{r=R}$).

At a higher level of the program,
   an amoeba search routine
   reiterates this integration for different trial values of $(s,T_0)$,
   seeking to minimise the difference between the obtained $(R,M_\mathrm{d})$
   and target values chosen by the user.
Usually these targets are set to collect a sequence of solutions
   of different $R$ but equal $M_\mathrm{d}$,
   i.e. describing galaxies of identical mass composition
   but different compactness.

Once the $(s,T_0)$ values are known for a desired stationary model,
   the stellar pressure profile (\ref{eq.jeans.stationary})
   is integrated inwards from infinity and tabulated finely.
Trial values are chosen for the complex frequency of perturbation, $\omega$.
The stationary model's ODEs are integrated again
   simultaneously with the coupled ODEs for the perturbed variables:
   (\ref{eq.mass.perturbed}),
   (\ref{eq.dark.mass.1}),
   (\ref{eq.dark.mass.2}),
   (\ref{eq.dark.momentum.1}),
   (\ref{eq.dark.momentum.2}),
   (\ref{eq.star.displacement}),
   (\ref{eq.star.mass}),
   (\ref{eq.jeans.stationary})
   and
   (\ref{eq.jeans.lambda}).
When the halo surface is reached ($r\rightarrow R$)
   the outer boundary condition (\ref{eq.obc}) is tested there,
   to assess the closeness of $\omega$ to an eigenvalue.
}
Using a complex test score,
\begin{equation}
	Z=Z_{\rm R}+iZ_{\rm I}=(\lambda_{P_{\rm d}}-\Lambda_{P_{\rm d}})
	\ ,
\label{eq.score}
\end{equation}
   a root-finding routine
   refines the $\omega$ trial values until $Z\rightarrow0$.
To seek eigenfrequencies on the real and imaginary axes of $\omega$,
   a bisection root-finder is sufficient.
To find eigenfrequencies elsewhere in the complex plane,
   a two-dimensional amoeba searcher is unleashed on the $Z=Z(\omega)$ landscape.
Due to symmetries of the ODEs in their $\omega^2$ factors,
   the eigenfrequencies can be:
   purely imaginary (\S\ref{s.stability});
   purely real (\S\ref{s.oscillations});
   or occuring in complex conjugate pairs (\S\ref{s.complex}).

If numerical integration traverses many orders of magnitude in $r$
   then unfortunately there are instances
   in which one or more of the $\lambda$ variables
   overflows or underflows the computer variables.
In practice, numerical integration towards
   the outer boundary produces diverging $|\lambda_{P_\mathrm{d}}|$,
   if $\omega$ is poorly chosen.
This must be tackled through a practical numerical trick.
One can pick the worst offending variable
   ($\lambda_j=\lambda_{_\mathrm{R}}+i\lambda_{_\mathrm{I}}$)
   and assign its magnitude to a new dependent variable,
   $\xi\equiv(\lambda_{_\mathrm{R}}^2+\lambda_{_\mathrm{I}}^2)^{1/2}$.
The code tracks the logarithmic profile of this quantity,
   by numerically integrating an extra ODE simultaneously with the other ODEs:
\begin{equation}
	{{{\mathrm d}\ln\xi}\over{{\mathrm d}r}}
	=
	{1\over{\lambda_{_\mathrm{R}}^2+\lambda_{_\mathrm{I}}^2}}\left({
		\lambda_{_\mathrm{R}}
			{{{\mathrm d}\lambda_{_\mathrm{R}}}\over{{\mathrm d}r}}
		+
		\lambda_{_\mathrm{I}}
			{{{\mathrm d}\lambda_{_\mathrm{I}}}\over{{\mathrm d}r}}
	}\right)
	\ .
\label{eq.autonormalisation}
\end{equation}
All the $\lambda$ variables can then be expressed in a locally normalised form,
   $\widehat\lambda_k\equiv\lambda_k/\xi$.
If each corresponding ODE ($k$) is abbreviated as a function
   ${\mathrm d}\lambda_k/{\mathrm d}r\equiv h_k(\lambda_1,\lambda_2,...)$
   then
\begin{eqnarray}
	{{\mathrm d}\over{{\mathrm d}r}}\widehat\lambda_k
	&\hspace{-3mm}=\hspace{-3mm}&
{\ifthenelse{\isundefined{\rainbow}}{}{\color{MidnightBlue}}%
	{{h_k(\lambda_1,\lambda_2,...)}\over{\xi}}
	-{{\lambda_k}\over\xi}\,{{{\mathrm d}\ln\xi}\over{{\mathrm d}r}}
	}
	\nonumber\\
	&\hspace{-3mm}=\hspace{-3mm}&
	h_k(\widehat\lambda_1,\widehat\lambda_2,...)
	-\widehat\lambda_k\,{{{\mathrm d}\ln\xi}\over{{\mathrm d}r}}
	\ .
\label{eq.norm.ode}
\end{eqnarray}
{\ifthenelse{\isundefined{\rainbow}}{}{\color{MidnightBlue}}%
The simplification $h_k(\lambda)/\xi=h_k(\widehat\lambda)$
   depends on the linearity of the perturbations.
}
Evolving (\ref{eq.autonormalisation})
   for $\xi=\xi(r)$
   in conjunction with the ODEs of the modified form
   (\ref{eq.norm.ode})
   keeps all the $\widehat\lambda$ variables normalised
{\ifthenelse{\isundefined{\rainbow}}{}{\color{MidnightBlue}}%
   to moderate values
   that do not overflow the floating-point representations.
Keeping the running values small also avoids
   driving the numerical integrators into stiff or unreliable domains.
Near the outer boundary, where $\lambda_{P_\mathrm{d}}$
   tends to become large,
   the running values of $\widehat\lambda_{P_\mathrm{d}}$
   and $\widehat\Lambda_{P_\mathrm{d}}$
   can be kept on the order of unity or smaller.
This prevents numerical noise in
   the sign of a score
   $\widehat{Z}\equiv Z/\xi$
   that is based on (\ref{eq.score}).
}
For presentational purposes,
   the $\widehat{\lambda}$ profiles can be rescaled by $\xi$
   afterwards in post-processing.

For brevity and computational ease,
   we choose physical units such that
   $G=1$, along with the stellar half-light radius $R_\mathrm{e}=1$
   and the stellar density there $\rho_\mathrm{e}=1$.
Results can be scaled into metric or astronomical units
   by inserting fiducial values of these constants for a known galaxy.

Families of models will be labelled by their total stellar mass fractions,
   ${\mathcal F}_\bigstar\equiv M_\bigstar/(M_{_{\rm d}}+M_\bigstar)$.
This paper presents galaxy models
   for a variety of choices of ${\mathcal F}_\bigstar$.
A fiducial ``cosmic'' baryon fraction $\approx0.163$
   is defined as in \cite{saxton2008} and \cite{saxton2010}.
For poorly understood reasons,
   observed densities of baryons
   \citep[e.g.][]{persic1992,bell2003,xue2008,anderson2010,mcgaugh2010}
   and dark matter
   \citep{karachentsev2012}
   differ from cosmic mean estimates (except in rich galaxy clusters).
{\ifthenelse{\isundefined{\rainbow}}{}{\color{MidnightBlue}}%
There may be an unseen ambient sea of both materials,
   free from self-bound objects.
Matching the abundances of observed galaxies
   to simulated CDM halo distributions
   implies baryon fractions $\la0.04$
   depending on galaxy mass
   \citep[e.g.][]{guo2010,papastergis2012}.
This may mean that galaxies lost most of their primordial gas endowments,
   or it may indicate that real halo populations differ from CDM.
Cosmological initial conditions are beyond the scope of the present analyses,
   so ${\mathcal F}_\bigstar$ is for now a free parameter.
Nonetheless \S\ref{s.stability}
   shows that spherical SIDM galaxies
   have stability limits that may influence
   the emergent distribution of ${\mathcal F}_\bigstar$.

In collisionless CDM, the outer halo density
   drops gradually without a well-defined boundary.
These simulated objects
   are conventionally circumscribed by the ``virial radius,''
   containing some multiple of the cosmic critical density, e.g.  
   $\rho_\mathrm{v}=100\rho_\mathrm{crit}
	\approx1.4\times10^4\,m_\odot\,\mathrm{kpc}^{-3}$.
If the natural density and radius scaling of SIDM haloes
   is not too different from CDM then we can estimate
   upper limits to the halo radii of represenative galaxies,
   $R^3\la{3M_\bigstar/4\pi\rho_\mathrm{v}{\mathcal F}_\bigstar}$.
For the example of M87,
   $R/R_\mathrm{e}\la30/{\mathcal F}_\bigstar^{1/3}$;
for M32,
   $R/R_\mathrm{e}\la200/{\mathcal F}_\bigstar^{1/3}$;
for a red nugget,
   $R/R_\mathrm{e}\la120/{\mathcal F}_\bigstar^{1/3}$;
and for the Fornax dSph
   $R/R_\mathrm{e}\la10/{\mathcal F}_\bigstar^{1/3}$
   \citep[data:][]{walker2009}.
It is however unclear whether this rule-of-thumb applies to SIDM,
   which warrants new cosmic structure formation calculations.
}%

\begin{figure}
\begin{center}
\includegraphics[width=84mm]{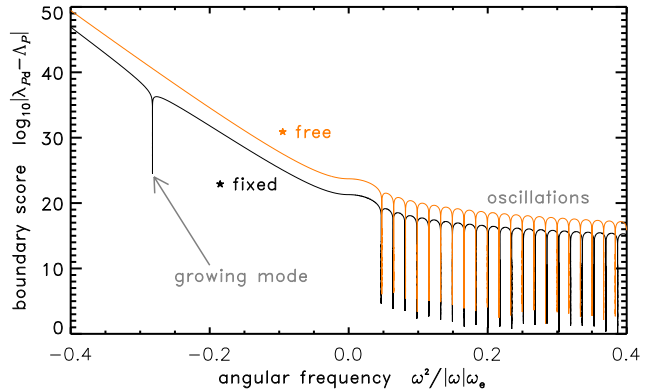}
\end{center}
\caption{%
Plot of the difference between calculated and ideal values of
  $\lambda_{P_\mathrm{d}}$ at the outer boundary,
   for trial $\omega$ values.
The horizontal axis combines real and imaginary $\omega$ adjacently.
Sharp dips reveal eigenfrequencies.
For the black curve, the stars were a fixed background potential.
For the orange curve, the stars were free to oscillate.
The ``growing mode'' dip
   disappears when stars and dark matter are gravitationally coupled
   and both free to move.
This galaxy has parameters
  $F=8$,
  ${\mathcal F}_\bigstar\approx0.163$,
  $R=100\,R_{\mathrm e}$.
}
\label{fig.coupling}
\end{figure}

\section{Results}
\label{s.results}

\subsection{stability properties}
\label{s.stability}

Classical arguments based on secular conditions
   suggested that $F>6$ polytropes
   should be unstable to runaway expansion or collapse
   \citep[e.g.][pp~51--53]{ritter1878,emden1907,chandrasekhar1939}.
This tendency manifests itself 
   (in the present formulation)
   as a conjugate pair of modes with imaginary $\omega$ values.
These ``collapse modes'' or ``growing modes''
   involve no oscillatory motion,
   only exponential growth or decay.
In practice such modes are found by searching
   along the imaginary axis of the $\omega$ Argand plane.

Next, a polytropic halo can be tested
   in a static stellar background
   (acting as a fixed potential, with
   $\lambda_{r_\bigstar}=\lambda_{P_\bigstar}=\widetilde{m}_\bigstar=0$).
The stability properties can be compared to the generic system
   where both the stars and dark matter are enabled to move radially.
Figure~\ref{fig.coupling}
   illustrates modes of a particular model with $F=8$,
   comparing the results if the stars are fixed or free.
With stars held fixed, a collapse mode appears.
With the stars enabled to move,
   the collapse mode vanishes from the model shown.
This stabilisation is typical of galaxies
   where the halo extends farther than $R>10R_\mathrm{e}$.

{\ifthenelse{\isundefined{\rainbow}}{}{\color{MidnightBlue}}%
The stabilisation can be explained qualitatively as follows.
SIDM fluid carries pressure waves,
   and $F>6$ modes are collapsible because
   pressure may not rise strongly enough
   to counteract compressive perturbations.
Stars however tend to persist on their orbits
   unless tidally disturbed,
   and do not conduct true sound waves.
They deafen the propagation of overdensities,
   and exert a local damping and stabilising effect.
For many galaxy models this suffices to confer global stability.
However, as we shall see below,
    in extreme cases the stabilisation is insufficient (\S\ref{s.slow})
    or a new type of instability emerges
    (\S\ref{s.fast}; \S\ref{s.superfast}).
}%

\begin{figure*}
\begin{center}
\includegraphics[width=18cm]{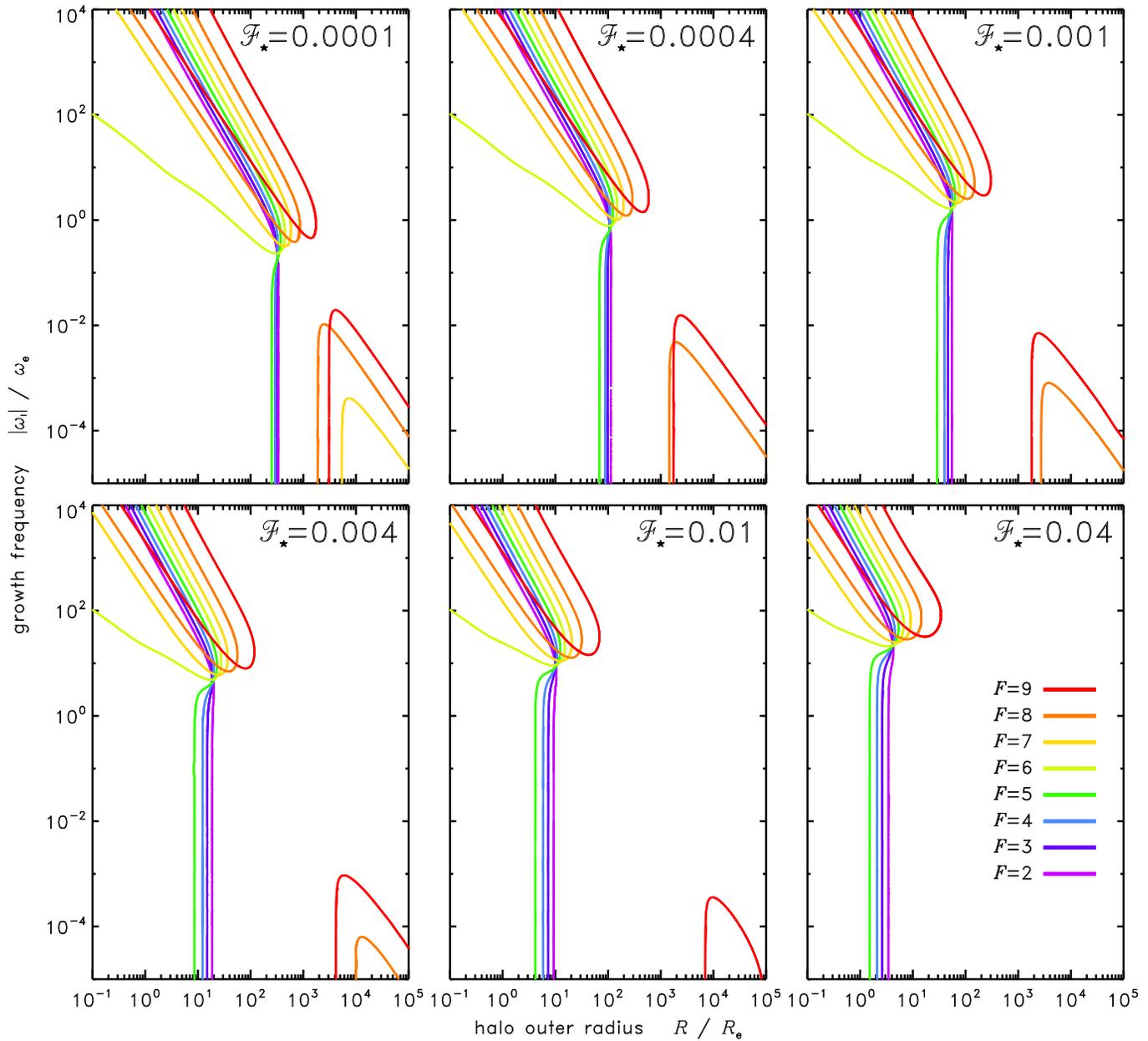}
\end{center}
\caption{Frequencies of the halo collapse modes
   as a function of the outer radius of the halo,
   for models dominated by dark mass.
Curves are coloured in spectral order
   to show the SIDM thermal degrees of freedom:
   from violet ($F=2$) to red ($F=9$).
Panels portray different values of the global stellar mass fraction.
As ${\mathcal F}_\bigstar$ is increased,
  ``fast'' modes rise in frequency,
  ``slow'' modes (bottom-right curves)
  recede to affect only the most radially extended haloes.
}
\label{fig.limits.R1}
\end{figure*}

The rest of this paper 
   assumes that the dark matter and stars are gravitationally coupled
   and both respond to radial perturbations.
Exhaustive numerical surveys have located the collapse modes
   as functions of:
   the equation of state parameter ($F$),
   the radial extent ($R$) of the halo,
   and the global richness of stars (${\mathcal F}_\bigstar$).
Figure~\ref{fig.limits.R1}
   illustrates the $|\omega|$ collapse frequencies as functions of 
   $R$ and $F$ (colours),
   for models where dark matter dominates
   (small ${\mathcal F}_\bigstar$, in respective panels).
Figure~\ref{fig.limits.R3}
   compares the modes
   when DM is present in the ``cosmic'' abundance,
   and a case that is DM-poor.
Realistic haloes could have radii of tens or hundreds of kpc,
   perhaps a range $10\la{R/R_\mathrm{e}}\la10^3$ depending on the galaxy.
Ultra-dense and ultra-diffuse cases are also plotted
   to clarify some theoretically notable modes.
Two distinct types of modes occur in particular domains.

\subsubsection{slow growing modes}
\label{s.slow}

If $R$ is large (a very diffuse halo)
   and $F>6$ then an instability occurs.
This is a reemergence of the instability
   of classic one-component polytropes with $F>6$.
In these cases the stars are unable to suppress instability
   because the stellar density is too low in the outer halo.
This happens over a larger domain of $R$
   if ${\mathcal F}_\bigstar$ is lower
   (dark matter dominates stars).
As $R$ is increased, the mode turns on suddenly:
   for $R$ below this threshold,
   a galaxy is safe;
   for slightly larger $R$ the mode operates on its shortest timescale;
   for very large $R$ the eigenfrequency diminishes
   ($|\omega|\sim R^{-3/2}$).
In a survey of configurations with $7\le F\le9$
   in the range $0.0001\le{\mathcal F}_\bigstar\le0.64$,
   the frequency scale of these ``slow'' growth modes
   is lower than the effective stellar frequency scale,
   $|\omega|<0.02\omega_\mathrm{e}$.
In other words, the $e$-folding duration of the growth / collapse
   would last several tens of orbits for most stars.
The slow mode straddles the accoustic frequency scale:
   always $|\omega|\la8\Omega_\mathrm{d}$ for $F=9$ haloes
   of the largest radii considered,
   but $|\omega|\la1\Omega_\mathrm{d}$ more usually.
The $|\omega|/\Omega_\mathrm{d}$ ratios
   are lower for $F=8$ than $F=9$,
   and lowest for $F=7$.

\begin{figure}
\begin{center}
\includegraphics[width=84mm]{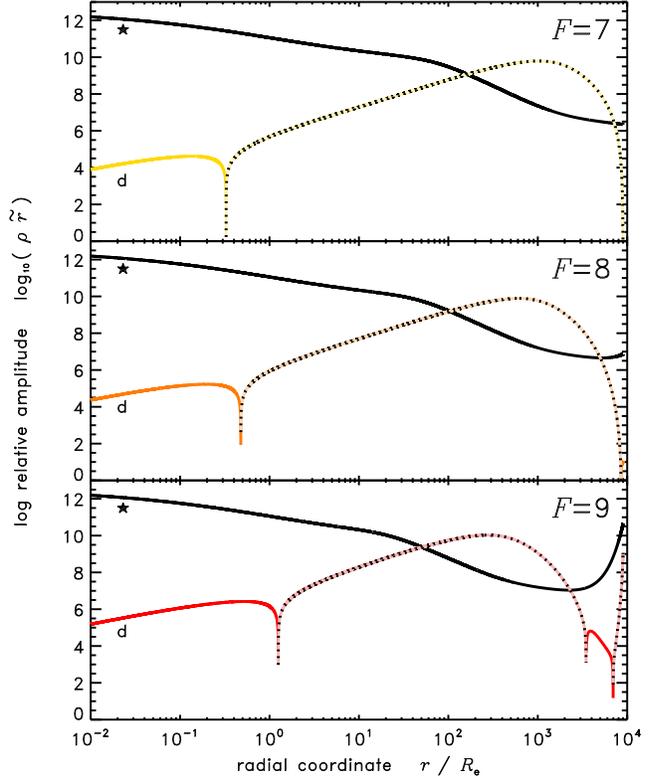}
\end{center}
\caption{%
Radial profiles of the slow mode
   in DM-rich galaxies (${\mathcal F}_\bigstar=0.0001$)
   with large halo ($R=9000R_{\rm e}$)
   and $F=7, 8, 9$.
The horizontal axis is radius ($r$);
   the vertical axis is a density-weighted displacement or flow,
   $\log_{10}(\rho\widetilde{r})$.
The stellar profile is black (``$\bigstar$'');
   dark matter is coloured
   (``d'', using the palette in Figure~\ref{fig.limits.R1}).
Curves are solid or dot-faded in the parts where the sign
   is positive or negative respectively.
Downward spikes are nodes at rest.
The vertical normalisation is arbitrary.
}
\label{fig.collapse.slow}
\end{figure}

The perturbed mass flux density through a sphere of given radius is
   $\widetilde{\rho v} = i\omega\rho_0\tilde{r}$,
   so the quantity
   $\rho_0\tilde{r}$
   is a useful expression of the radial flows of stars and dark matter.
Figure~\ref{fig.collapse.slow}
   depicts the radial flow structure during ``slow mode'' evolution.
Amplitudes for stars tend to be highest in the centre and decline radially.
The dark flow has its minimum in the centre.
The stars tend to move in the same direction everywhere.
Dark matter has alternating ingoing and outgoing layers,
   separated by nodes at rest.
In the interior, dark matter moves the same way as the stars.
Then there is a node 
   at some radius around $\ga0.1R_{\rm e}$ to $\sim10R_{\rm e}$
   (farther out for larger $F$).
Outside this node, the dark flow is in antiphase with the stars.
Some models have further nodes and flow reversals
   at larger radii in the outer halo
   (e.g. two far outer nodes in the $F=9$ example
   in Figure~\ref{fig.collapse.slow}).
The stability analyses are symmetric under time reversal,
   so it is impossible to say whether a particular sign implies
   collapse or expulsion of matter.
The slow mode could imply
   contraction of all stars and the inner dark matter
   while an external dark layer escapes.
It could also mean radial expansion of stars and inner halo,
   while dark outskirts converge inwards.

\subsubsection{fast growing modes}
\label{s.fast}

If $R$ is sufficiently small
   then one or two growing modes appear.
They occur for all $F$ choices.
The affected models have hot and dense haloes,
   with much of the dark mass residing
   inside the half-light radius.
These ``fast'' modes occur at high frequencies:
   usually $|\omega|\gg\omega_\mathrm{e}$,
   (and often
   $|\omega|\gg10^2\,\omega_\mathrm{e}$
   for larger ${\mathcal F}_\bigstar$ cases).
Where there are two ``fast'' eigenvalues,
   the uppermost is quick compared to the acoustic crossing of the halo
   ($|\omega|>\Omega_\mathrm{d}$, in many models by a few dex).
The lower eigenvalue is faster than $\Omega_\mathrm{d}$ if $F\ge7$,
   but can be slower than $\Omega_\mathrm{d}$ for $F<7$.

\begin{figure}
\begin{center}
\includegraphics[width=84mm]{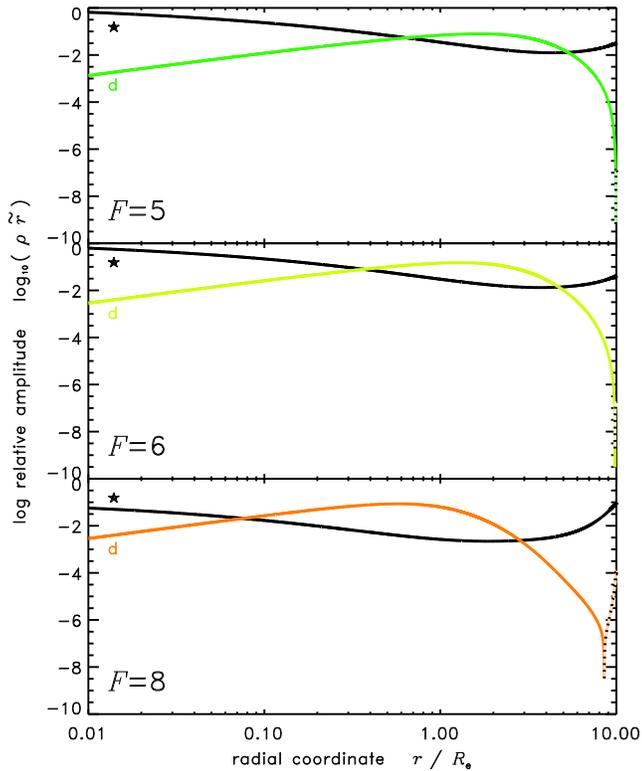}
\end{center}
\caption{%
Radial wave profiles of the lower-$\omega$ fast growing mode
   in galaxy models with
   stellar mass fraction ${\mathcal F}_\bigstar=0.01$,
   halo radius $R=10R_{\rm e}$ and $F=5, 6, 8$.
The horizontal axis is radius ($r$);
   the vertical axis is
   $\log_{10}(\rho\widetilde{r})$.
Dark matter is coloured (``d'');
   the stellar profile is black (``$\bigstar$'').
Curves are solid or fade-dotted in the parts where the sign
   is positive or negative respectively.
}
\label{fig.collapse.fast1}
\end{figure}

\begin{figure}
\begin{center}
\includegraphics[width=84mm]{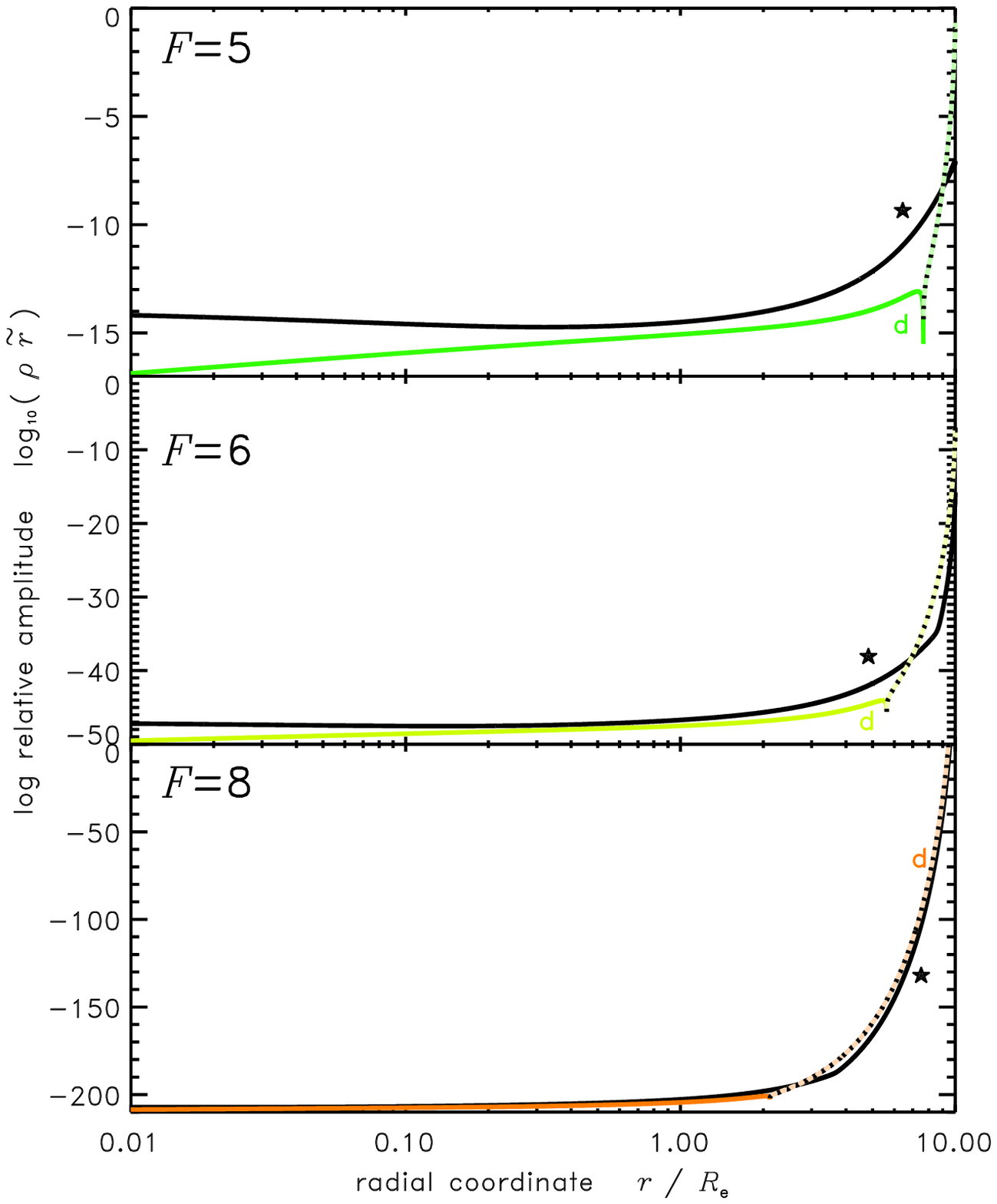}
\end{center}
\caption{%
Wave profiles of the higher-$\omega$ fast growing mode
   corresponding to Figure~\ref{fig.collapse.fast1}.
The displacement amplitudes rise steeply with radius;
   amplitudes are most significant at the surface
   and negligible in central regions.
For larger $F$ the ampitude rise is greater
   and the DM node is at smaller radius.
}
\label{fig.collapse.fast2}
\end{figure}

Figure~\ref{fig.collapse.fast1}
   depicts the mass flows due to the lower-$\omega$ fast-mode
   for illustrative models with halo radius $R=10R_{\rm e}$
   and DM enriched over ``cosmic'' composition.
In some ways these profiles resemble the waveforms of slow modes:
   the stellar amplitude is high in the interior;
   the dark matter amplitude is peaked farther out.
The dark matter has one node near the outer boundary,
   well outside the half-light radius of stars.
Except in the outskirts,
   the dark mass is flowing in the same direction as the stars.
For the corresponding higher-$|\omega|$ fast modes
   (Figure~\ref{fig.collapse.fast2})
   there is again just one node in the dark matter,
   but both constituents show a dramatic rise in amplitude
   towards the outer boundary.
The rise is steeper for larger $|\omega|$ and larger $F$.
These fast modes act mainly in the diffuse outskirts:
   rapidly driving the stars and dark matter in opposite directions.
The core is essentially unaffected.

{\ifthenelse{\isundefined{\rainbow}}{}{\color{MidnightBlue}}%
Since the fast instability occurs for all $F$,
   its main causes must be traits of the stellar profile.
We might suspect the stellar cusp and its coldness
    ($\mathrm{d}\sigma_\bigstar/\mathrm{d}r>0$ at the origin)
   however tests show fast modes afflicting
   non-cuspy models with a \cite{plummer1911} profile of $\rho_\bigstar$.
If the exponential stellar fringe is a factor,
   the fast modes might differ in tidally truncated models
   \citep[generalising][]{king1966}
   which will be analysed in a future paper.
In the present model,
   the index of the stellar gravitational field
   ($\mathrm{d}\ln{g_\bigstar}/\mathrm{d}\ln{r}$)
   curves significantly near $r\sim10R_\mathrm{e}$
   but settles to the asymptote $-2$ farther out.
If the halo surface occurs at $r\la30R_\mathrm{e}$
   then the oscillatory restoring force varies in form during the cycle.
Fast instabilities may result from the inadequacy of restoring forces
   for haloes that truncate in this region.
}

\begin{figure}
\begin{center}
\includegraphics[width=84mm]{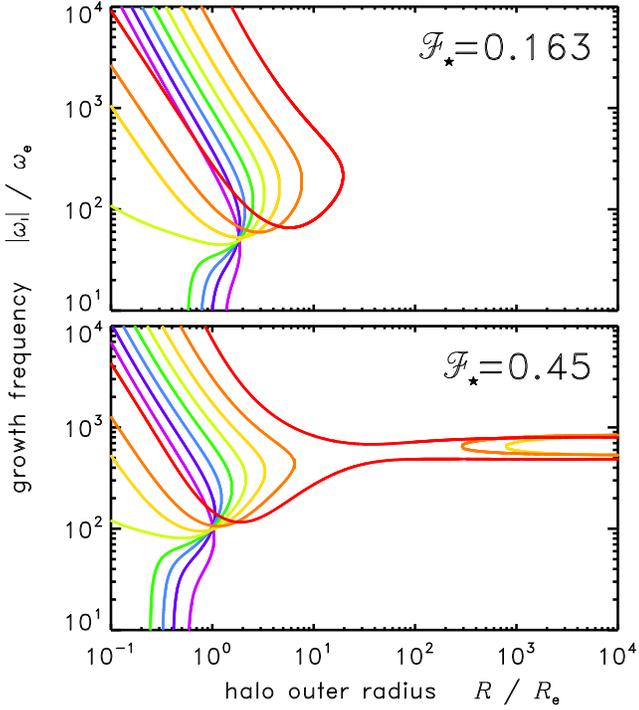}
\end{center}
\caption{%
Frequencies of the halo collapse modes
   as a function halo radius $R$
   for galaxies with cosmic abundances (top)
   and DM-depleted cases (bottom).
The halo $F$ values are coloured as in Figure~\ref{fig.limits.R1}.
As ${\mathcal F}_\bigstar$ is increased,
   the ``fast'' instability mode
   begins to affect radially large $F>6$ models
   as well as the compact and $F\le6$ models.
These modes become ubiquitous for all $R$,
   if $F>6$ and stellar mass is sufficiently dominant.
}
\label{fig.limits.R3}
\end{figure}

\subsubsection{safe configurations}
\label{s.safe}

\begin{figure}
\begin{center}
\includegraphics[width=84mm]{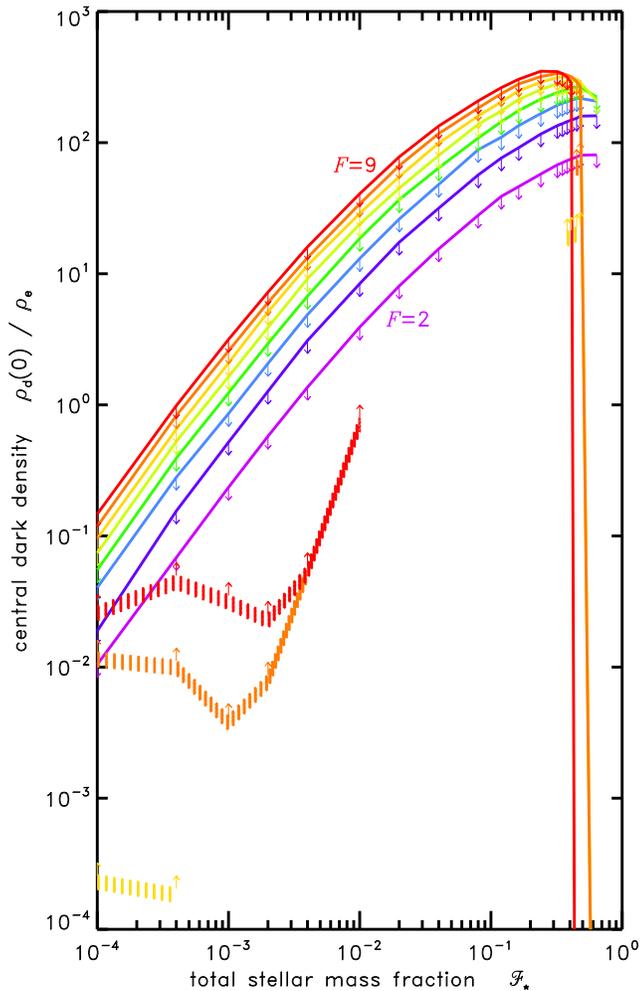}
\end{center}
\caption{%
Central density of dark matter at the limits of fast and slow instabilities
   (solid and dashed lines respectively).
The domain between the curves is safe.
The horizontal axis is the global stellar mass fraction
   (${\mathcal F}_\bigstar$).
}
\label{fig.limits.dens}
\end{figure}

\begin{figure}
\begin{center}
\includegraphics[width=84mm]{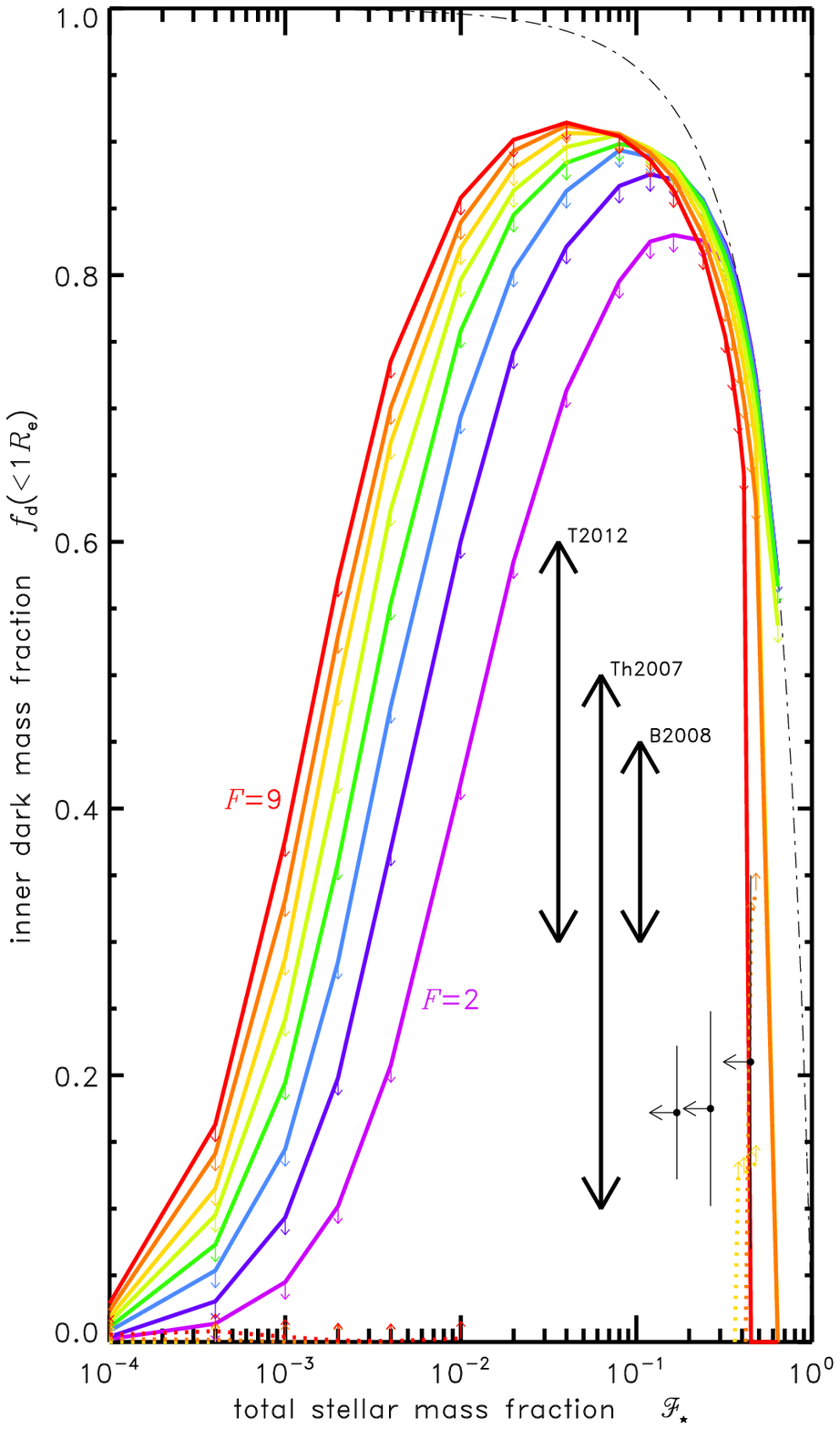}
\end{center}
\caption{%
Stability limits on the dark mass fraction
   within $1R_\mathrm{e}$
   corresponding to states in Figure~\ref{fig.limits.dens}.
The grey dot-dashed line indicates the maximum
   dark mass fraction possible for given ${\mathcal F}_\bigstar$.
Estimates for individual elliptical galaxies
    are black dots with ${\mathcal F}_\bigstar$ upper limits
    \citep{thomas2005,murphy2011,norris2012}.
Thick arrows show $f_\mathrm{d}(1R_\mathrm{e})$ ranges
    from galaxy samples
    \citep{bolton2008,thomas2007,tortora2012}
    drawn here at arbitrary ${\mathcal F}_\bigstar$.
}
\label{fig.limits.conc1}
\end{figure}

Fast modes and slow modes are absent
   throughout a special intermediate range of $R$.
In these cases 
   the interaction between stars and the halo suppresses
   the collapse mode
   (e.g. in the ``coupled'' case of Figure~\ref{fig.coupling}).
These haloes with $F>6$
   are stabilised by the mingling of collisionless stars,
   contrary to na\"{i}ve textbook expectations.
The three radial domains ---
   defined by the ``fast'' instability, stable zone, and ``slow'' instability
   ---
   imply limits on the compactness of the stable endpoints of galaxy evolution.
The fast growing modes mean that models
   are unstable unless $R$ exceeds some lower limit
   (which is a slowly varing function of $F$ and ${\mathcal F}_\bigstar$).
The slow mode only afflicts $F>6$ cases,
   and it applies an upper limit on $R$.
This upper limit is a sensitive function of ${\mathcal F}_\bigstar$.
For models with $F>6$ that are poor in stars
   (${\mathcal F}_\bigstar\la0.001$),
   the slow- and fast-mode domains
   leave only a narrow band of safe $R$.
This safe band becomes wider as ${\mathcal F}_\bigstar$ increases.
A numerical survey indicates that the slow mode (upper limit)
   vanishes for sufficiently high ${\mathcal F}_\bigstar$.
(It persists till
	${\mathcal F}_\bigstar\la0.0004$ for $F=7$,
	${\mathcal F}_\bigstar\la0.004$ for $F=8$,
	and
	${\mathcal F}_\bigstar\la0.01$ for $F=9$.)
From this point till ${\mathcal F}_\bigstar\la0.4$,
   galaxies with arbitrarily wide outer radii $R$ are stable.
Figure~\ref{fig.limits.dens}
   illustrates the stability limits
   on the central dark density
   ($\rho_{\rm d}|_{r=0}$)
   as a function of ${\mathcal F}_\bigstar$.
Figure~\ref{fig.limits.conc1}
   shows the corresponding limits on
   the dark mass fraction within the half-light radius
   ($f_{\rm d}(<1R_{\rm e})$).
The latter quantity can be measured
   through gravitational lensing
   and measurements of kinematic tracers.
Significantly, it appears that a SIDM halo
   around an isolated de~Vaucouleurs type galaxy
   is only stable if the dark mass within $1R_{\rm e}$
   is less than $\approx90\%$ of the total,
   or less depending on ${\mathcal F}_\bigstar$ and $F$.

Mass modelling of observed elliptical galaxies
   often indicates $f_{\rm d}(<1R_{\rm e})$ values
   in the range of a few tens of percent
   \citep{loewenstein1999,ferreras2005,thomas2005,thomas2007,cappellari2006,
		bolton2008,tortora2009,weijmans2009,
		saxton2010,grillo2010,memola2011,murphy2011,
		norris2012,tortora2012,wegner2012}.
If these galaxies are stable within polytropic haloes,
   we could broadly infer $0.005\la{\mathcal F}_\bigstar\la0.3$
   independently of $F$.
Observational estimates of the stellar mass are consistent with this range.
No large nearby galaxy is yet known to challenge
   the density limit of the fast instability.
A few exceptional galaxies appear DM-poor at large radii
   \citep{mendez2001,romanowsky2003,salinas2012}.
If their impoverishment is not an illusion due to orbital anisotropies,
   then we might explain them in terms of
   runaway halo loss via either of the instabilities
   (\S\ref{s.dm.poor}).

\subsubsection{fast modes in DM-poor systems}
\label{s.superfast}

Entering the domain where stellar mass is globally dominant
   (Figure~\ref{fig.limits.R3})
   a new fast mode appears.
Beyond some threshold
   (${\mathcal F}_\bigstar\ga0.4$)
   an island of the fast modes begins to affect large-$R$ haloes
   with $F>6$.
This implies that DM-poor systems have a maximum stable radius,
   just like the DM-rich systems that are subject to slow modes.
The $\omega$ values of these fast modes
   are slowly varying functions of $R$
   (horizontal curves in bottom panel, Figure~\ref{fig.limits.R3})
   which implies they arise from
   the stars rather than the DM crossing time.

As ${\mathcal F}_\bigstar$ is increased further,
   the fast mode affects $F>6$ haloes with {\em all} possible $R$ values,
   and then there are no stable configurations with $F>6$.
For example when ${\mathcal F}_\bigstar=0.64$
   there are no stable models with $F=8$ or $F=9$.
The fast modes that afflict these DM-poor models
   are found to have steeply rising waveforms
   (mainly acting at the halo surface)
   resembling those in Figure~\ref{fig.collapse.fast2}.
Results were not sought beyond ${\mathcal F}_\bigstar=0.64$,
   because the system of equations becomes stiff
   in ways that make the $\omega$ root-finding slow and arduous.

\begin{figure}
\begin{center}
\includegraphics[width=84mm]{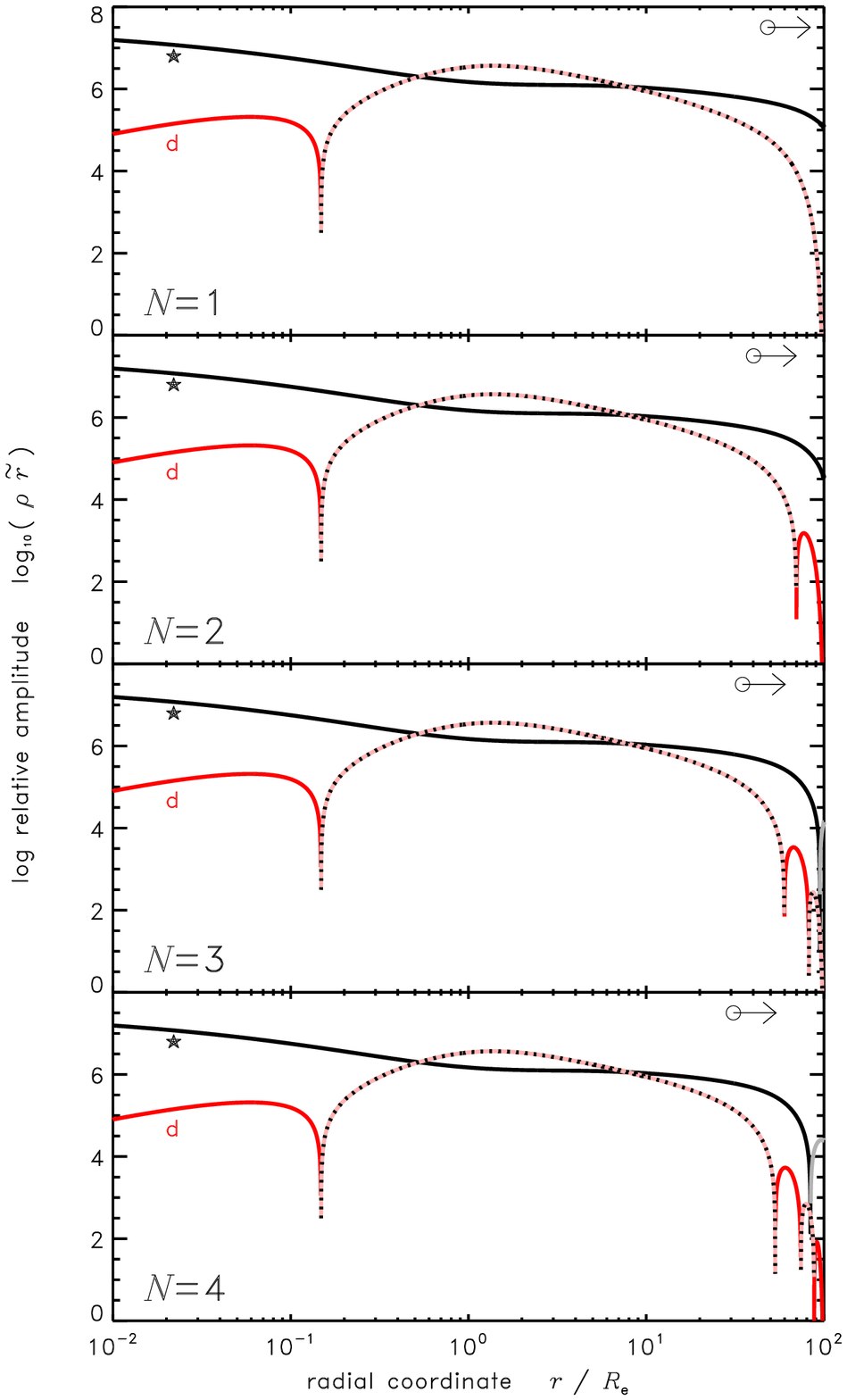}
\end{center}
\caption{Radial wave profiles of the first four oscillatory modes
   in a galaxy model with $F=9$ halo of radius $100R_{\rm e}$,
   and
   ${\mathcal F}_\bigstar\approx0.163$.
The vertical axis is
   $\log_{10}(\rho\widetilde{r})$.
Dark matter is coloured;
   the stellar profile is black.
Curves are solid or faded in the parts where the sign
   is positive or negative respectively.
Arrows indicate the radius beyond which the keplerian frequency
   is less than the radial oscillation frequency ($\kappa<\omega$).
}
\label{fig.harmonics}
\end{figure}

\subsection{neutrally stable oscillations}
\label{s.oscillations}
   
As in single-fluid asteroseismology,
   \citep[e.g.][]{eddington1918,cox1980,pijpers2006,aerts2010}
   each polytropic halo exhibits an infinite sequence of radial pulsation modes
   at discrete real values of $\omega$.
These radial modes are neutrally stable:
   neither growing nor decaying.
This is to be expected, as the present analysis is purely adiabatic:
   omitting any dissipation or mixing processes
   (e.g. shocks in the dark matter;
   violent relaxation of the stars;
   chaotic migration of stars).

\subsubsection{flow profiles of standing waves}
\label{s.flow}

Figure~\ref{fig.harmonics}
   illustrates examples of eigenfunction waveforms
   expressed in terms of oscillating mass flux density,
   ($\rho \tilde{r}$)
   for the dark matter (coloured curves) and stars (black curves).
For the stars, flow amplitudes
   are higher in central regions and decrease with radius.
For the dark matter, the peak amplitudes occur at intermediate radii.
Such profiles are generic to the radial pulsations of all models
   with the $(F,{\mathcal F}_\bigstar,R)$ values inspected so far.
Near the centre, the dark matter and stars oscillate in phase,
   but in outer regions they can oscillate in anti-phase.

For each constituent there are nodes at specific radii,
   where the matter remains locally at rest,
   and where there is a local reversal in the flow direction.
The number of nodes of the dark matter oscillations
   is a harmonic number, $N\ge0$.
Values of $N$ uniquely label the eigenmodes.
As $N$ increments,
   additional nodes appear in the outer fringe of the halo
   (e.g. between $50R_{\rm e}$ and $100R_{\rm e}$
   for the $N=1,\ldots4$ modes in Figure~\ref{fig.harmonics}).
Most of the nodes appear far enough out that
   the keplerian orbital frequency is locally
   less than the pulsation frequency:  $\kappa<\omega$
   (marked with arrows in Figure~\ref{fig.harmonics}).
{\ifthenelse{\isundefined{\rainbow}}{}{\color{MidnightBlue}}%
If wave amplitudes grew to the nonlinear regime,
   some tidal shocking and orbital migration of stars could occur.
Such dissipative effects scale as
   $\propto\rho\mathbf{v}\cdot\mathbf{g}$
   which is of non-linear order $\sim\varepsilon^2$,
   while the radial displacements are of order
   $\sim\varepsilon$.
Inwards/outwards pulsational displacements are an immediate effect,
   but orbital heating (and net outward expansion)
   could become significant gradually as more cycles elapse.
}

\begin{figure*}
\begin{center}
 \includegraphics[width=160mm]{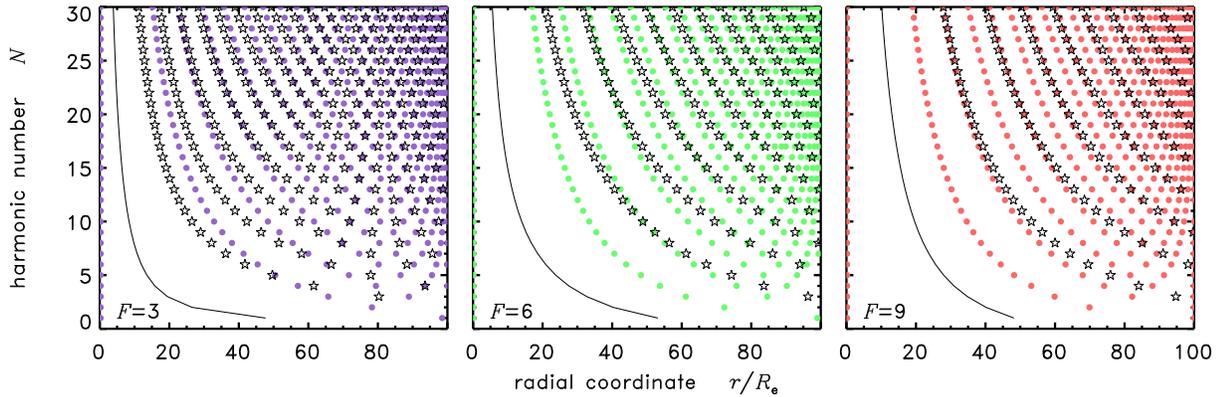}
\end{center}
\caption{%
Node locations of dark matter ($\bullet$ coloured)
   and stars ({\bf\ding{73}} symbols)
   for the $N\le30$ radial pulsation modes
   of galaxy models with
   ${\mathcal F}_\bigstar=0.163$,
   $R=100R_{\rm e}$
   and $F=3, 6, 9$ (left, middle and right panels respectively).
The black curve marks the radius where $\kappa=\omega$.
}
\label{fig.nodes.F}
\end{figure*}

Depending on $N$ and the galaxy model,
   a number of nodes can also occur in the stellar profile
   ($N_\bigstar$).
For a galaxy with $F=9$,
   ${\mathcal F}_\bigstar=0.163$ and $R=100R_{\rm e}$,
   the node numbers are 
   $(N,N_\bigstar)=(1,0), (2,0), (3,1), (4,1), (5,1), (6,2)$
   $\ldots (9,3), \ldots (18,5)$
   and so on.
Figure~\ref{fig.nodes.F} shows the dark and stellar node positions
   for the first thirty oscillatory modes of this model,
   and for equivalent models where $F=3$ and $F=6$.
It appears that $N_\bigstar$ never decreases as $N$ increases,
   but there is no explicit link between these numbers.
The sequence of $(N,N_\bigstar)$ and their radii
   provide a fingerprint of a particular galaxy,
   and must be obtained numerically.

Figure~\ref{fig.nodes.F}
   also shows that the dark nodes
   are more crowded near the outer boundary ($R$)
   in halo models with larger $F$.
This is a consequence of
   slower wave propagation in the fringes,
   since small-cored (large $F$) models
   have proportionally lower $\sigma$ profiles in the outskirts.
Stellar nodes are more sublty affected by $F$.
For models with small $F$, the stellar nodes
   are more numerous and reach smaller radii for given $N$.
Haloes with large $F$ tend to have fewer stellar nodes,
   and they are distributed farther out.
Varying the stellar mass fraction has slight effects
   on the positions of the nodes in the outskirts.
Across a wide domain $0.002\le{\mathcal F}_\bigstar\le0.48$
   (but with fixed $N$, $F$ and $R$),
   the stellar nodes do not shift significantly.
The outer nodes of the dark matter appear more affected:
   they occur at radii a few percent farther out if
   ${\mathcal F}_\bigstar$ is very small.

\begin{figure*}
\begin{center}
\begin{tabular}{ccc}
 \includegraphics[width=160mm]{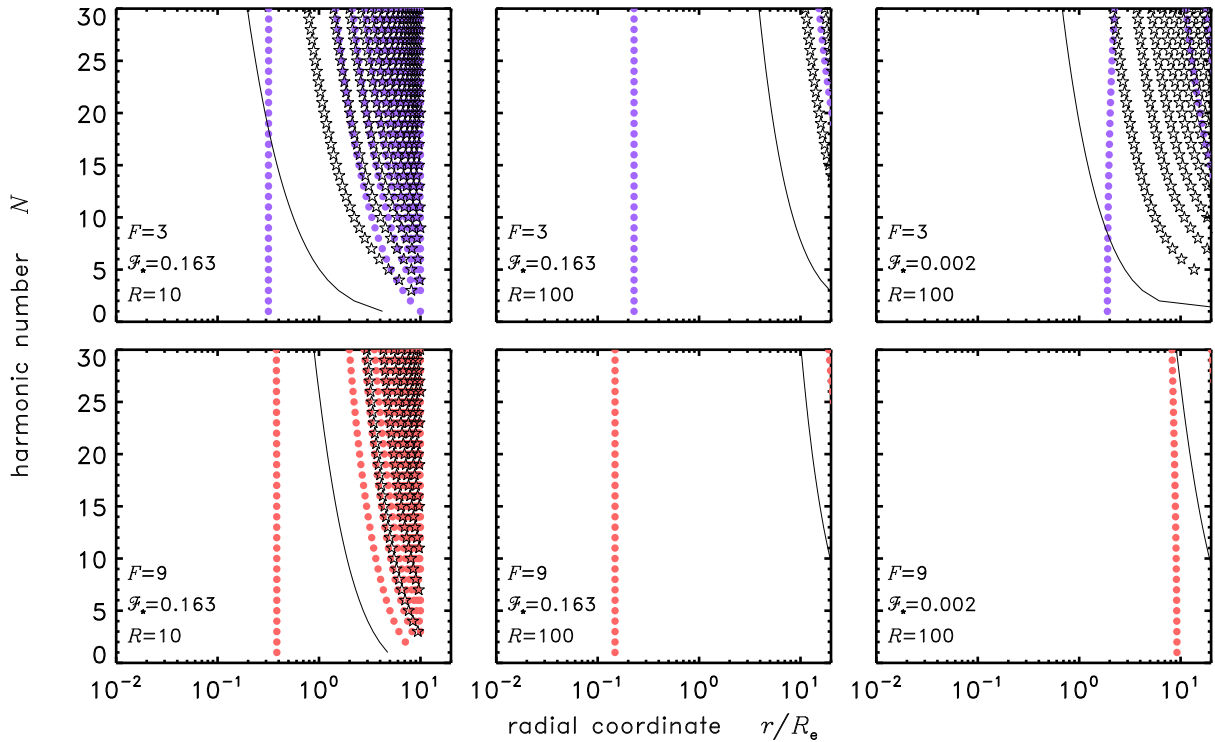}
\end{tabular}
\end{center}
\caption{%
Node locations for $N<30$ pulsations
   in the inner regions of model galaxies with
   $F=3$ (top row) and $F=9$ (bottom row)
   and various halo radii and stellar mass fractions
   ($R=10, 100R_{\rm e}$
   and ${\mathcal F}_\bigstar=0.002, 0.163$ as annotated).
The innermost dark node occurs at a smaller radius
   if the stellar mass is relatively large (big ${\mathcal F}_\bigstar$).
Symbols and the $\kappa=\omega$ line are marked as in Figure~\ref{fig.nodes.F}.
}
\label{fig.nodes.RY}
\end{figure*}

For waves that have at least one DM node ($N\geq1$),
   the radius of the innermost node
   varies more widely than any other node:
   from $\sim0.1R_{\rm e}$ to $\approx10R_{\rm e}$
   in the domain of $0.002\le{\mathcal F}_\bigstar\le0.48$,
   $2\leq F\leq9$ and $10R_\mathrm{e}\leq R\leq100R_\mathrm{e}$.
Its location varies only sublty with harmonic number $N$.
Figure~\ref{fig.nodes.RY} depicts the detailed distribution of nodes
   in the inner, astronomically observable regions ($r<20R_{\rm e}$)
   of galaxies with $F=3$ and $F=9$
   where the halo radius is $R/R_{\rm e}=10, 100$
   and the DM content is rich or ``cosmic''
   (${\mathcal F}_\bigstar=0.002, 0.163$).
All else being equal,
   the innermost dark node tends to occur
   at smaller radii if the galaxy is DM-poor (large ${\mathcal F}_\bigstar$),
   and farther out if the galaxy is DM-rich (small ${\mathcal F}_\bigstar$).
For DM-rich haloes, a larger radius $R$ shifts the inner dark node
   to larger radii.
For some DM-poorer haloes, larger $R$ shifts the inner dark node inwards.
Comparison of the $F=3,9$ cases in Figure~\ref{fig.nodes.RY}
   also shows how the lower-$F$ models have more stellar nodes
   at radii small enough that ripples might be observable
   in the profiles of starlight and stellar velocities.

\subsubsection{frequency spectrum}

\begin{figure*}
\begin{center}
\begin{tabular}{ccc}
 \includegraphics[width=164mm]{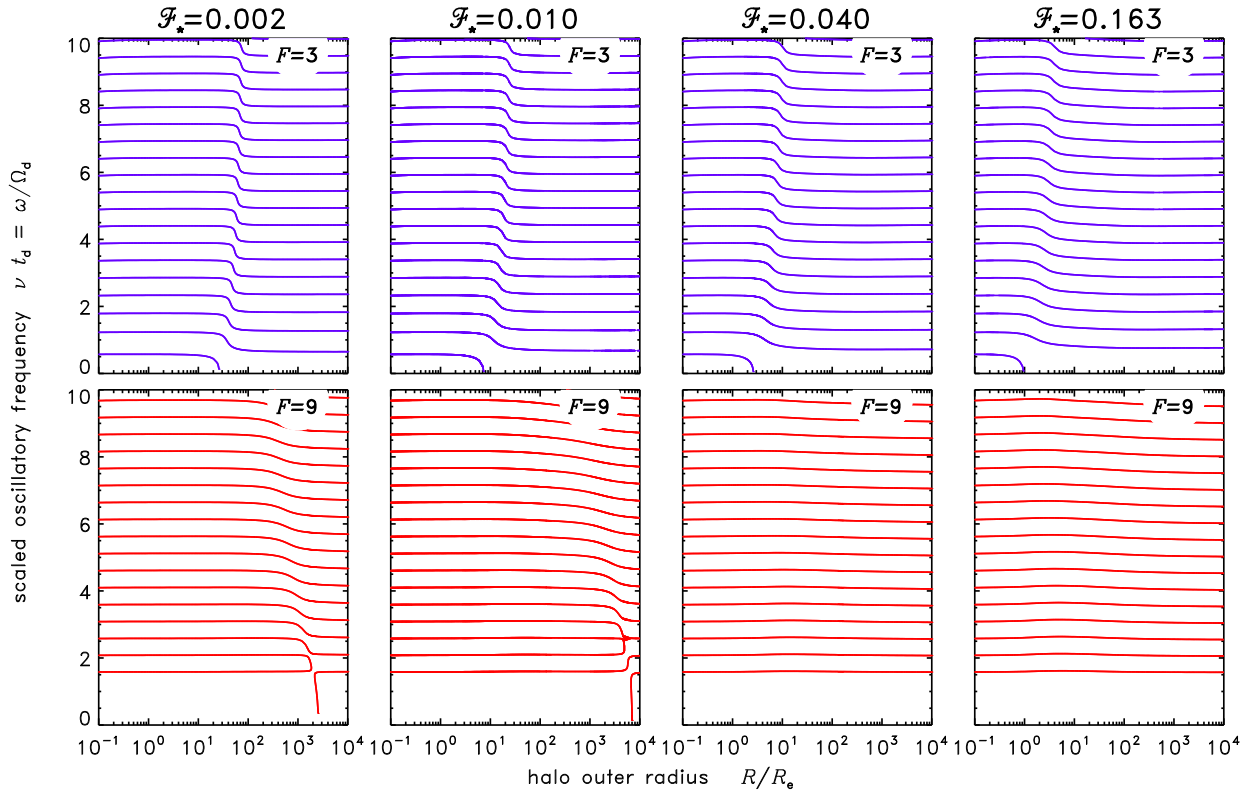}
\end{tabular}
\end{center}
\caption{%
Oscillatory spectra for galaxies
   for various choices of the halo radius $R$ (horizontal axes).
Frequencies (vertical axis) are scaled by
   the acoustic crossing time of the halo.
Panels show cases
   where the halo has $F=3$ (top row) and $F=9$ (bottom row).
The stellar mass fraction is ${\mathcal F}_\bigstar=0.002, 0.01, 0.04, 0.163$
   in the columns from left to right respectively.
The frequencies are nearly linearly spaced,
   except in intervals of $R$ where one of the lower modes vanishes.
}
\label{fig.sizes}
\label{fig.modes}
\end{figure*}

Figure~\ref{fig.modes}
   illustrates the oscillatory spectra
   for various baryon fractions(${\mathcal F}_\bigstar$)
   and thermal degrees of freedom $F=3$ or $F=9$.
Among these and other results,
   it is usually found that
   most eigenfrequencies $\omega$
   are approximately linearly spaced in relation to $N/t_{\rm d}$
   (where $t_{\rm d}$ is the acoustic crossing time).
Typically $\Delta\nu\approx0.5\,/t_{\mathrm d}$
   (or $\Delta\omega\approx0.5\,\Omega_\mathrm{d}$)
   independently of $(F,{\mathcal F}_\bigstar,R)$.
If the mode spacing of a real galaxy halo were measurable,
   this information would provide a diagnostic of the halo radius
   via its crossing time (see Appendix~\ref{s.stationary}).
The confirmed detection or inference of the {\em lowest} mode
   would constrain $(F,{\mathcal F}_\bigstar,R)$ more tightly.

There are domains where the lowest mode deviates from the linear spectrum.
For galaxies with sufficiently compact halo (small $R$)
   there exists a fundamental oscillation
   that has no dark matter nodes in the waveform ($N=0$).
Keeping the same $(F,{\mathcal F}_\bigstar)$
   but increasing the halo radius ($R$),
   there is a narrow range of $R$
   where the fundamental frequency detaches from the rest of the spectrum,
   and drops to zero frequency ($\omega_0\rightarrow 0$).
For radially larger haloes,
   the lowest mode is the first overtone ($N=1$).
For $F<6$ the fundamental mode drops out 
   somewhere from a few $0.1R_{\rm e}$ to a few $10^2R_{\rm e}$.
The dropout threshold $R$ is greater in DM-rich systems
   (small ${\mathcal F}_\bigstar$).
For haloes with $F\ge6$ the $N=0$ dropout occurs below $R<0.1R_{\rm e}$,
   so that all such galaxies of realistic size
   lack a fundamental mode (if $F\ge6$).
In the $(F,{\mathcal F}_\bigstar)=(9,0.002)$
   and $(9,0.010)$ cases of Figure~\ref{fig.modes},
   we see a further dropout of the $N=1$ mode when $R$ is very large.
The dropout range in $R$
   is not in any obvious way related
   to the onset of fast and slow instabilities.
Apart from influencing the dropout radius,
   the stellar mass fraction (${\mathcal F}_\bigstar$)
   has little effect on the shape of the oscillatory spectrum.

{\ifthenelse{\isundefined{\rainbow}}{}{\color{MidnightBlue}}%
For halo sizes that are realistic for elliptical galaxies
   ($10^1\la R/R_{\rm e}\la10^3$)
   the fundamental mode is nonexistant.
If so then all oscillatory modes have at least one dark node.
The {\em innermost} dark node
   turns out to occur at radii sufficiently small
   that there may be consequences for the visible galaxy.
Since this node location is nearly independent of $N$,
   it seems possible that dark eigenmodes could excite each other
   via disturbances in the DM close to this region.

The stellar nodes are stationary points
   in the radial displacement of the stars.  
The mean radial velocities would be zero at these points of rest,
   and {\em opposite} signs on either side of each node.
Rippling overdensities and underdensities of stars
   would appear between consecutive nodes.
Such density variations might be observable
   if any stellar nodes occur within $r\la10R_\mathrm{e}$,
   where the surface brightness is practically measurable.
In the present model calculations,
   stellar nodes tend to occur farther out than the dark nodes,
   but some stellar nodes can appear at smaller radii if $F$ is small
   or if the oscillations are of high order.
For observable ripples to appear,
   the order of the overtone needs to be large ($N\gg30$).
}%

\subsection{mixed complex modes}
\label{s.complex}

As emphasised in \S\ref{s.numerical}, the $\lambda$ ODEs
   only depend on the perturbation frequency via $\omega^2$ terms.
For a one-component polytrope
   the eigenfrequencies should be purely real or purely imaginary.
The two-component galaxy model however 
   is an innately more complicated (double) oscillator,
   due to the $\tilde{g}_\pm$ gravitational couplings.
Even if $\lambda_{r_{\rm d}}$ and $\lambda_{r_\bigstar}$
   are in phase at the inner boundary,
   they may be out of phase at other radii
   due to the different respective wave speeds.
It is possible for some eigenfrequencies to occur
   off the real and imaginary axes,
   in conjugate pairs
	($\omega=\omegaR+i\omegaI$, $\omega^*=\omegaR-i\omegaI$).

In models where such modes do occur,
   they coexist with the ordinary oscillations and growth modes.
Finding them requires extra effort to explore the $\omega$ plane in 2D.
The landscape of outer boundary scores,
   $Z\equiv\lambda_{P_{\rm d}}-\Lambda_{P_{\rm d}}$,
   exhibits steep gradients ($\partial Z/\partial\omega$),
   which easily confounds ``amoeba'' or ``steepest descent'' solvers.
It is more practical to map $Z$ on a fine rectangular grid.
The desired solutions appear as contour rings
   or spikes on surface plots of $\log|Z|$
   \citep[e.g. as in analyses of shock instabilities][]{
		toth1993,saxton1998,saxton2001}
   or by tracing the intersections of
   $Z_{_\mathrm{R}}=0$ and $Z_{_\mathrm{I}}=0$ contours.
When an eigenfrequency is approximately located,
   an amoeba solver can be unleashed in that vicinity
   to refine the $\omega$ solution.

\begin{figure*}
\begin{center}
\includegraphics[width=140mm]{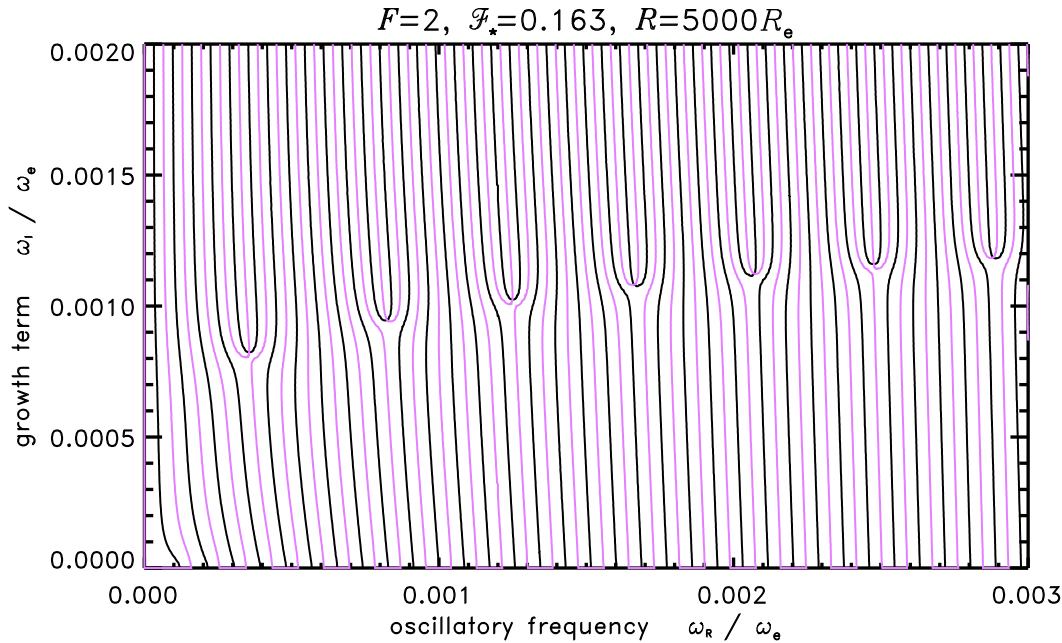}
\end{center}
\caption{%
Map of the boundary condition score,
   $Z\equiv(\lambda_{P_{\rm d}}-\Lambda_P)$,
   in part of the $\omega$ complex plane,
   for a large-$R$ galaxy model.
Black contours show where the real part is zero;
   violet contours show where the imaginary part is zero.
The intersections of black and violet contours are complex eigenfrequencies
   (seven such knots appear in this plot).
}
\label{fig.complex.map}
\end{figure*}

Figure~\ref{fig.complex.map}
   locates the first few complex modes of a representative system.
Complex modes occur in
   a minority of the galaxy models investigated so far.
These tend to have radially large enough
   ($R\ga10^3R_{\rm e}$)
   that the crossing times are similar
   for the dark matter and stars,
   $t_\bigstar\approx t_{\rm d}$
   (see curves in Figure~\ref{fig.steady.time} in the appendices).
Such huge halo radii (Mpc scale) are astronomically unrealistic,
   and so the mathematical existence of complex modes
   may be inconsequential to galaxies.
Complex modes may be of more interest
   if they manifest in other analogous astrophysical systems:
   e.g. a star cluster possessing a gaseous atmosphere;
   or collisionless particles bound inside a fluid star.

\begin{figure}
\begin{center}
\includegraphics[width=84mm]{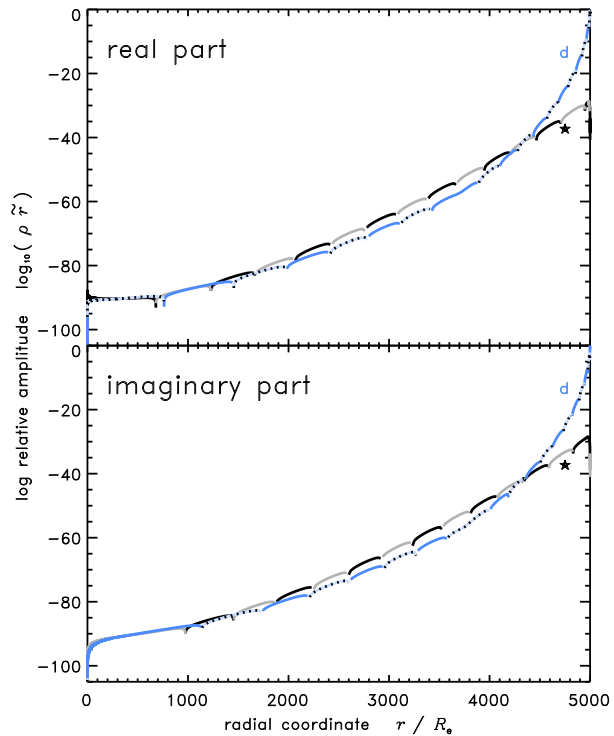}
\end{center}
\caption{Log-amplitudes of density-weighted displacements
   ($\rho\tilde{r}$)
   for stars (black) and dark matter (coloured)
   in the lowest complex mode of the galaxy model with
   $F=4$, $R=5000R_{\rm e}$ and ${\mathcal F}_\bigstar=0.5$.
The eigenfrequency $\omega=(0.740+2.751i)\times10^{-3}\omega_\mathrm{e}$.
Top and bottom panels show the real and imaginary parts.
}
\label{fig.complex.wave}
\end{figure}

Like the fast growth modes, the complex modes 
   have wave amplitudes that are small in the interior
   and rise faster than a power-law towards the halo's surface.
Figure~\ref{fig.complex.wave} shows the profile
   of the mass displacement ($\rho\tilde{r}$)
   corresponding to the lowest-$|\omega|$ complex mode
   of a model with $F=4$.
As $\omega$ is complex
   the eigenfunction has real and imaginary parts too.
Radial nodes are numerous for all complex modes,
   because $|\omega|\gg\Omega_\mathrm{d}$.
The innermost node for the dark matter motions is around
   $r\approx0.13R_{\rm e}$,
   similar to the innermost node of
   this galaxy's real-$\omega$ pulsations.
The only complex modes found so far are in $F=2, 3, 4$ haloes,
   where the relatively high DM densities in the outskirts
   enable significant gravitational coupling
   between the components there.
For fixed $({\mathcal F}_\bigstar,R)$
   larger $F$ causes the complex $\omega$ eigenfrequencies
   to be spaced farther apart.

Whether the high-frequency oscillation ($\omegaR$)
   combined with growth ($\omegaI$)
   is a fatal instability depends on non-linear effects.
High-frequency rippling
   ought to induce violent relaxation and orbital mixing of stars.
This would introduce non-linear damping,
   and may prevent the mode from growing to destructive amplitudes.
Proving such damping effects would require higher-order perturbative analyses,
   including work terms,
   ($\propto\rho{\mathbf v}\cdot{\mathbf g}\sim\varepsilon^2$).

\section{Discussion \& Implications}
\label{s.discussion}

\subsection{reversibility}

The collapse modes of the cores of giant stars
   are considered to cause some classes of supernovae.
Are galaxy haloes susceptible to dark explosions
   if they possess the analogous collapse mode?
The analogy should not be overstretched.
Stellar core collapse is irreversible
   due to the accompanying nuclear reactions (and loss of neutrinos),
   and subsequently due to an emerging event horizon
   \citep[e.g.][]{goldreich1980}.
If these microphysical processes have counterparts
   in the SIDM domain
   then the convulsions of a galaxy-scale halo
   could sometimes have irreversible results.
Otherwise, the dark matter collapse modes
   could produce repeated, reversible, non-linear bouncing.
At least four phenomena could incur some irreversibility:
\begin{enumerate}
\item
If the dark particles can self-scatter,
   then their self-annihilation and decay channels
   may also be non-negligible in some astrophysically attainable conditions
   \citep[e.g.][]{buckley2010,feng2010b}.
High densities of SIDM might react
   to yield neutrinos, photons or other escaping byproducts.
This is an ``unknown unknown,''
   beyond the scope of the present paper.
\item
Shock-heating of the SIDM
   would raise the halo entropy
   \citep[unless the dark shocks are reversible,
	as in a scalar field model of][]{peebles2000}.
A shocked halo will eventually settle
   with a different radius $R$
   but with $(F,{\mathcal F}_\bigstar)$ unchanged.
\item
Tidal shocking and mixing of the stellar orbits
   could alter the orbit distribution irreversibly.
This could in principle 
   affect the evolution of luminosity verses radius relations
   of elliptical galaxies.
Early-type galaxies in violently oscillating haloes
   might evolve lower stellar densities
   via tidal exchanges driving stellar diffusion radially.
\item
Strong compressions of the inner halo
   could perhaps form a central black hole,
   or feed it significantly.
The possibility of forming supermassive black holes (SMBH)
   via parsec-scale Jeans-unstable hydrodynamic ``dark gulping''
   was discussed by \cite{saxton2008}.
The masses involved are only an inner portion of a halo,
   and any collapse requires $F>6$.
Non-linear effects and long free-fall times would prevent
   the halo from vanishing into the horizon {\em entirely}.
\end{enumerate}

\cite{ostriker2000}, \cite{hennawi2002} and \cite{balberg2002a}
   also considered growing black holes from SIDM,
   by a mechanism that depends on a gravothermal catastrophe
   in thermally conductive, weakly scattering haloes
   (not magnetic SIDM, nor scalar field SIDM).
\cite{balberg2002a} predict plausible submassive black hole seeds
   in galaxy haloes, starting from a cored initial condition.
\cite{ostriker2000} and \cite{hennawi2002}
   set initial conditions with a thermal inversion
   in a central density cusp
   (similar to the hypothetical cusps of collisionless CDM).
From this they claimed tight upper limits on SIDM interactivity.
Either way,
   gravothermal generation of a SMBH requires the SIDM scattering crossection
   to be in a special range that ensures kpc-scale mean-free-paths.
If this requirement is unmet,
   collapse need not be imminent within a Hubble time
	\citep{balberg2002b,ahn2005,koda2011}.

\subsection{contrast to collisionless models}

To calculate the discrete and continuous modes
   of entirely collisionless galaxies
   is a complicated task,
   due to the higher dimensionality and freedom
   available to phase-space distribution functions;
   or due to a lack of closure of the Jeans equations
   for large-amplitude perturbations.
Elaborate matrix formulations are normally used
   \citep{kalnajs1977,polyachenko1981,palmer1987,weinberg1991,saha1991}.
These involve choosing orthogonal basis functions,
   and practical applications require infinte matrices
   to be truncated approximately.
More elegant methods exist for special systems
   \citep[e.g.][]{sobouti1984,sobouti1985,sobouti1986}.

Despite the complication of a live halo as a second mass component,
   the present formulation is tractable because of
   several fortunate conditions.
The existence of the dark halo's boundary conditions
   constrains the freedom of the collisionless stellar component indirectly
   via their gravitational coupling and the $\Lambda_\mathrm{P}$ target.
The finite halo radius ensures that the collective modes
   occur at discrete values of $\omega$,
   rather than on continua.
The assumption of small purely radial perturbations
   enables the zero-torque condition (\ref{eq.torque}).
(Analysis of non-radial modes would require a different approach.)

Much of the literature concerning the modes of collisionless galaxies
   has concentrated on the presence of
   ``radial orbit instabilities''
   which are non-spherical ($l=2$) unstable modes
   of systems where stellar orbits are radially biased
   \citep{polyachenko1981,palmer1987,weinberg1991,bertin1994,trenti2006}.
In theory, non-spherical deformations could arise spontaneously
   in some objects that are stable against radial perturbations.
Though this paper focusses on radial modes,
   we can at least say that radially unstable systems
   are likely to be unstable to non-radial motions too.
Persistent non-radial pulsations are a possibility,
   and they could alter the instantaneous ellipticities and asymmetries
   of galaxies and their haloes.
It is at this stage unknowable whether non-radial instabilities
   could destroy radially stable galaxies.
A general non-radial analysis would be worthwhile,
   if another closure condition could be derived
   to replace the zero-torque assumption.

\subsection{pulsating galaxies}

It was long ago recognised that collisionless galaxy models,
   if built from superpositions of integrable orbits,
   are capable of persistent pulsations
   \citep[e.g.][]{louis1988,vandervoort2003}.
These require sharp gaps in the phase-space distributions of stars.
Unfortunately the gaps
   are difficult to resolve numerically in $N$-body simulations,
   due to the innate limits of mass discretisation.
Perhaps because of this systematic effect,
   the possibility of pulsating galaxies is rarely mentioned
   in simulations and observational research.
Orbital chaos and diffusion processes might
   eventually dampen these oscillations in real galaxies and star clusters.

In the present paper,
   it is shown that galaxies that combine stars with a fluid-like dark halo
   can sustain coupled oscillations,
   without fine requirements on the stellar orbit distributions.
A disturbed mass of SIDM would echo and jiggle
   with its own internal sound waves.
These {\em darkquakes} displace
   the visible swarm of stars in their orbits.
Reciprocally, the stars affect the dark waves,
   e.g. removing the instability of many $F>6$ models.
Concentric radial overdensities of apparently displaced stars,
   if observed in nature,
   could be a signature of galaxy pulsations.
If so then these would be an interesting diagnostic
   of the dark physics,
   enabling studies of {\em skotoseismology}
   --- a seismology of dark matter.
The density ripples of high-overtone darkquakes
   (with kiloparsec wavelengths)
   would in principle have subtle effects
   on the gravitational lensing profiles of galaxies.

``Shell galaxies'' are observed to have sharp stellar overdensities
   in bands encircling the main host galaxy
   \citep[e.g.][]{malin1980,quinn1984,hernquist1987,prieur1988,
	turnbull1999,wilkinson2000}.
These are conventionally thought to be
   phase-wrapped remnants of dwarf galaxies destroyed in minor mergers.
Stars from the disrupted secondary galaxy
   are strewn preferentially in certain orbits about the primary,
   and migrate till they occupy shells
   (rather than narrow streams).
This is an attractive model when the observed shells differ
   from the stellar populations of the host galaxy,
   and when shell radii interleave in steps between the left and right sides.
If however any case were found
   in which shell stars match the host populations exactly,
   or if any shell encircled the galaxy completely,
   then we might alternatively propose 
   that host stars were displaced radially by dark halo pulsations
   (of high enough order to reach radii $\la10R_\mathrm{e}$).
If the shell intervals appear smaller at larger radii,
   this would signify proximity to the halo's truncation edge, $R$.

Passive elliptical galaxies observed at high redshifts
   appear to have modern stellar masses ($\sim10^{11}m_\odot$)
   but contained within half-light radii
  ($R_\mathrm{e}\sim1$~kpc)
   much smaller than modern counterparts
	\citep{cimatti2004,toft2005,daddi2005,trujillo2006,
		longhetti2007,vandokkum2008,damjanov2009,
		vandesande2011,trujillo2011,toft2012,ferreras2012}.
A mysterious process drives these ``red nuggets''
   to inflate on Gyr timescales,
   but without appreciable addition of mass,
   nor any significant starbursts.
Expulsion of gas by active nuclei or stellar activity
   might conceivably alter the galaxy's potential
   \citep[e.g.][]{fan2008,damjanov2009}.
Gasless minor mergers
   might inflate the fringes of the stellar profile
   \citep[e.g.][]{khochfar2006,naab2009}.
Now we might invoke an additional explanation, in terms of halo seismology.
For stable galaxies, large oscillations could be excited
   by flybys or disturbances beyond the visible stellar outskirts.
Stochastic excitations might be effective
   in cosmic environments of any density.
If higher overtone pulsations reach large enough amplitudes,
   the outer stars may suffer tidal shocking or orbital diffusion,
   leading to the observed radial expansion.
However if the initial conditions of the red nuggets
   already transgress the stability limits
   (Figures~\ref{fig.limits.dens} and \ref{fig.limits.conc1})
   then mass will redistribute spontaneously on timescales
   $\sim\omega_\mathrm{e}^{-1}$, even without external stimulation.
Compariing these theorised evolutionary channels
   to empirical redshift-growth trends
   is worth a dedicated study elsewhere.

{\ifthenelse{\isundefined{\rainbow}}{}{\color{MidnightBlue}}%
\subsection{extremes of dark matter content}
\label{s.dm.rich}
}%

The stability limits on spheroidal galaxies with SIDM haloes
   are loosest for stellar mass fractions around
   ${\mathcal F}_\bigstar\sim0.04$.
The constraints are tighter for systems 
   with very small or large ${\mathcal F}_\bigstar$
   (DM-rich and DM-poor extremes).
Thus the theory predicts
   limits on the stable end-points of galaxy evolution.
These depend on intrinsic structural properties
   rather than traits passively inherited
   from cosmological initial conditions and halo merger histories.

\label{s.dm.poor}
The strongest finding is the exclusion of
   DM impoverished galaxies (${\mathcal F}_\bigstar\ga0.5$)
   when the halo has a high heat capacity ($F>6$).
If such a system were born,
   the ubiquitous fast modes
   (\S\ref{s.superfast})
   would destroy it by one process or another.
The flux profiles imply dramatic evolution in the outskirts:
   perhaps runaway loss of the halo
   (till only a bare star cluster remains).
Runaway accretion of dark matter is another possibility
   (if an external reservoir is available)
   until ${\mathcal F}_\bigstar$ is lowered
   to a stable configuration.
The observed non-existence of stationary isolated galaxy-scale systems
   with ${\mathcal F}_\bigstar\ga0.5$
   is consistent with SIDM predictions.

At the opposite extreme in DM richness,
   observed dwarf spheroidal galaxies (dSph)
   are aged, quiescent stellar systems
   with low luminosities
   $10^3\la L/L_\odot\la10^7$
   and dark masses $>10^7\,m_\odot$
   within $1R_\mathrm{e}$
   \citep[e.g.][]{mateo1998,gilmore2007,strigari2008,walker2009,
		walker2011,amorisco2012,jardel2012}.
For the Milky Way satellites,
   $0.80<f_\mathrm{d}(<1R_\mathrm{e})\la0.999$
   which implies ${\mathcal F}_\bigstar<0.2$
   for the brightest
   and ${\mathcal F}_\bigstar<10^{-2}$
   for ultra-faint cases.
It appears that dSph occupy the upper-left, low-${\mathcal F}_\bigstar$ part
   of Figures~\ref{fig.limits.dens} and \ref{fig.limits.conc1},
   where their stability is problematic.
Their stellar density profiles are
   not cuspy like (\ref{eq.ps.density})
   but rather cored like a \cite{plummer1911} model.
Preliminary calculations with this profile
   reveal fast and slow instabilities resembling those reported above.
Cuspiness is unimportant;
   the effect of tidally truncated outskirts is less clear.
It may also be significant that {\em satellite} galaxies
   are not isolated (\S\ref{s.boundary}).
Immersion of dSph in the host galaxy's halo
   would reduce the unstable domains.
The analysis of galaxies with truncated \cite{king1966} profiles
   and a stabilising external pressure
   awaits a future paper.

\hspace{1mm}
{\ifthenelse{\isundefined{\rainbow}}{}{\color{MidnightBlue}}%
\subsection{alternative boundaries and asymmetries}
\label{s.boundary}
}%

A real halo need not be isolated in vacuum.
External medium confinement can stabilise polytropes
   \cite[e.g.][]{mcrea1957,bonnor1958,horedt1970,umemura1986}.
This effect may stabilise clustered galaxies and dwarf satellites
   that would be unstable in the field.
An ambient background of unbound dark matter
   could imply a warm halo surface ($\sigma_\mathrm{d}^2>0$).
If external DM accretes supersonically,
   signals cannot propagate upstream
   and the isolated halo model still holds.
If the contact is somehow subsonic
   then open boundary conditions apply instead of (\ref{eq.obc}):
   waves might escape rather than reflecting at the surface.
If the interface has a high contrast
   (like the solar atmosphere)
   then the vacuum boundary calculations
   might approximate $\omega$ well enough.
If the halo blends smoothly into its surroundings 
   then the oscillatory frequency intervals might decrease to the external Jeans frequency
   ($\sqrt{G\rho_\mathrm{x}}$).

If the outskirts are infinite and fluid-like
   but the asymptotic density is near-zero
   then the eigenmodes would merge into a continuum:
   a functional relation between
   $\omegaR$ and $\omegaI$.
This relation would be found by setting $\lambda$ values at infinity,
   integrating radially inwards
   and matching the inner boundary conditions
   (\ref{eq.ibc.r})--(\ref{eq.ibc.star2}).
This top-down approach is used in analyses of radiative shocks
   \citep[e.g.][]{saxton1997}.
The stellar profile (\ref{eq.ps.density})
   is radially infinite,
   though real galaxies may trucate at a tidal radius.
If so,
   an outer boundary condition on $\lambda_{P_\bigstar}$
   analogous to (\ref{eq.obc})
   would affect the modes.
If the stellar boundary is cold ($\sigma_\bigstar=0$)
   then stellar nodes would be bunched there,
   and ripples could appear more numerous than in \S\ref{s.flow}.
As long as either the SIDM or the stars are spatially finite,
   a discrete spectrum is expected.

In weak-SIDM models with effectively collisionless outskirts
   \citep[e.g.][]{yoshida2000b,vogelsberger2012,peter2012,rocha2012},
   the outer halo would
   act as part of the ``$\bigstar$'' component in this paper.
The fluid-like ``d'' component would
   reside in isolation within a reduced boundary radius $R$,
   shortening the crossing time and raising the pulsation frequencies.

Unless a galaxy is well isolated,
   its halo may receive cosmological infall of smaller objects.
If the outskirts are quasi-collisionless
   then these mergers may cause tidal disturbances
   that excite oscillations in the core.
If the outskirts are in a strong-SIDM condition,
   then weak shocks occur,
   the impactor can ablate and dissolve,
   and ripples may propagate
   (akin to raindrops in a pond).
Mergers and mixing raise
   the total halo entropy ($\propto M_\mathrm{d}\ln s$),
   which may alter the final halo mass-radius trends.

Non-radial effects
   could complicate the oscillations and stability limits.
Disc galaxies (which provide the best evidence of cores)
   require axisymmetric models
   to assess the role of rotation
   \citep[e.g.][]{hachisu1986}.
Tidal fields, cosmic filaments and mergers
   could excite non-radial pulsations directly.
Non-radial instabilities might destroy some configurations
   that are radially stable.
The directionally dependent crossing times of a triaxial halo
   would split the frequencies,
   giving a non-linear spectrum of $\omegaI$ modes
   with angular dependencies.
Since $F>6$ polytropes are less stable than the standard $F=3$ case,
   we might expect larger shape distortions during non-radial pulsation.
The greatest attainable transient triaxiality
   must depend on $F$ and the environment;
   the extent is unknowable without 3D simulations.
It would be interesting to see the results of strong-SIDM $F>6$
   shape investigations extending \cite{peter2012}.

\vspace{2mm}
\subsection{new thermostatistics, dark fluid or exotica?}

Tsallis thermostatistics
   generalise the Boltzmann entropy
   in an attempt to describe systems
   where long-range interactions (such as gravity) are influential
   \citep{tsallis1988,plastino1993,nunez2006,zavala2006,vignat2011}.
If this conjecture applies to self-bound bodies of dark matter,
   then the optimal spherical configurations are polytropes,
   even without short-range interactions
   or scatterings among dark particles.
The equilibrium solutions for spherical galaxies
   have the same density and potential profiles
   as for SIDM fluid polytropes.
\cite{saxton2010}
   fitted polytropic halo models to observed elliptical galaxies
   without distinguishing between SIDM and Tsallis scenarios.

However in their time-dependent and dynamical properties,
   the SIDM polytrope is not entirely equivalent to
   the Tsallis ``stellar polytropes''
   \citep[e.g.][]{chavanis2002}.
Fluids and stellar polytropes
   differ in their phase-space distributions,
   and their microscopic physics.
If collisionless,
   dark matter need not support or transmit pressure waves.
For a collisionless dark halo (even with a polytropic profile)
   the necessary analytic methods
   would involve phase-space distribution functions
   or moments in Jeans equations,
   with all the problems that accompany the treatment of the stars.
The complexities of two-species collisionless modelling
   are beyond the scope of the present paper.
Such models might not
   exhibit the oscillations and stability properties calculated above.
Observation of darkquakes would favour SIDM over the Tsallis interpretation.

Besides SIDM, various alternative DM particle theories
   may be seismically distinct.
If haloes consist of neutral fermions,
   then degeneracy pressure may explain the cores
   while the fringes are collisionless
   \citep[e.g.][]{munyaneza2005,nakajima2007}.
Other candidates include ultra-light particles
   with de~Broglie lengths reaching kpc scales
   \citep[e.g.][]{sin1994,ji1994,lee1996,hu2000,rindler2012}.
Perturbations and collisions of these ``fuzzy DM'' galaxies
   produce interference patterns
   \citep{gonzalez2011}.
Such exotic halo models deserve their own stability analyses,
   in order to predict observable signatures 
   distinguishing them from SIDM and CDM.

\section{Conclusions}
\label{s.conclusions}

{\ifthenelse{\isundefined{\rainbow}}{}{\color{MidnightBlue}}%
This paper formulates perturbative stability analyses
   for spheroidal self-gravitating systems comprising
   a fluid mingled with collisionless matter.
The present application describes of spheroidal galaxies
   comprising a halo of self-interating dark matter
   and an empirical de~Vaucouleurs distribution of stars
   (with isotropic orbits in equilibrium).
The halo is polytropic, adiabatic, non-singular and radially finite.
It is isolated, with a vacuum outer boundary.
The perturbations from equilibrium are radial but non-uniform.}%

All radially finite halo models
   exhibit a spectrum of natural modes of radial pulsation.
These modes are neutrally stable,
   at least in the absence of high-order dissipative effects
   and stellar orbital mixing.
The frequencies are almost linearly distributed.
The frequency interval scales inversely
   with the halo's accoustic crossing time,
   which depends on the halo radius.
The thermal degrees of freedom $F$
   influence the frequency gap below the lowest mode.
For haloes of plausible radius ($R\gg10R_{\rm e}$)
   and central density,
   all oscillations have a dark node
   within ten stellar half-light radii.
If detected via its rippling displacement of gas and stars,
   these modes would help diagnose the actual radial extent
   and total mass of the dark halo.   
Nodes in the stellar oscillation are more likely to appear
   as detectable ripples and velocity patterns
   if the harmonic number $N$ is large,
   if the halo is compact (small $R$),
   or if there are few thermal degrees of freedom (small $F$).

Gravitational coupling of a collisionless stellar distribution
   with a polytropic dark halo
   significantly affects the collective stability of the galaxy.
Stability depends on
   the dark heat capacity (via $F$),
   the stellar mass fraction (${\mathcal F}_\bigstar$),
   and the halo radius ($R$).
Elliptical galaxies are stable
   for plausible ${\mathcal F}_\bigstar$
   and radii ($R$) smaller than Mpc scale,
   even in cases where the classical pure polytrope would be unstable ($F>6$).
This includes 
   the range of dark matter equations of state, $7\la F\la10$,
   favoured by observed galaxy kinematics,
   cluster core sizes
   and black hole masses
   \citep{saxton2008,saxton2010}.
The dark matter fraction within the stellar half-light radius
   never exceeds
   ${f_{\rm d}(<1R_{\rm e})}\la0.9$
   for any stable de~Vaucouleurs type configuration.
A spheroid born in an unstable state
   would presumably evolve towards one of the stable end-points.

The stability limits are defined by the onset
   of three types of growing modes.
Slow growing modes
   occur if the halo is radially large and $F>6$.
Slow modes resemble non-cyclic versions of the pulsation waveforms:
   maximum stellar flows near the centre,
   and maximum dark flows at intermediate radii,
   with DM and stars moving in opposite directions in some regions.
Fast growing modes (quicker than orbital periods at $1R_{\rm e}$)
   afflict galaxies with very centrally concentrated dark matter
   (regardless of $F$).
Isolated systems that are rich in dark matter
   are only stable if the central density is quite low
   and the halo radius is huge.
If stars dominate the galaxy mass,
   haloes with $F>6$ may suffer the fast instability
   regardless of the radius.
Complex modes are a mixture of oscillation and exponential growth.
They exist in principle but they only seem to afflict
   haloes of unrealistically vast radii.
Fast growing modes and complex modes
   entail eruption or collapse of mass layers
   mainly in the fringes of the halo,
   leaving the interior less disturbed.

The identification of these oscillations and instabilities
   (perhaps via visible stellar displacements
   or ripples in gravitational lens profiles)
   would place informative constraints on fundamental dark matter physics,
   and distinguish SIDM from other candidates.
If proven, more detailed skotoseismic analyses
   would provide complementary probes
   the mass and energy distribution surrounding individual galaxies.
This model also suggests an alternative interpretive framework
   that may help to explain:
   the expansion of red nuggets;
   the possible loss of haloes from extreme galaxies;
   morphologies of some shell galaxies;
   and perhaps some of the drivers of early-type galaxy scaling relations.

\section*{Acknowledgments}

I thank I.~Trujillo and I.~Ferreras for the conversation
   that challenged me to undertake this study.
I also thank K.~Wu, R.~Soria, I.~Ferreras, M.~Mehdipour, Z.~Younsi and D.~Kawata
for discussions during the work,
   and the anonymous referee for rigorous suggestions
   leading to extensive improvements.


\appendix

\section{core scales}
\label{appendix.radii}

A spherical polytropic halo without
   other gravitating components
   is a classic Lane-Emden sphere
   \citep{lane1870,emden1907}.
The core can be defined by
   locations where the logarithmic index of density
   reaches certain values,
\begin{equation}
	\alpha_\mathrm{d} \equiv
	-{{\mathrm{d}\ln\rho_\mathrm{d}}\over{\mathrm{d}\ln{r}}}
	\ ,
\end{equation}
   with $\alpha_\mathrm{d}=1$ at radius $R_1$,
   and $\alpha_\mathrm{d}=2$ at $R_2$.
These occur in fixed ratios
   to the halo surface ($R$).
The index $R_3$ is important because beyond this radius
   an isolated polytropic SIDM halo
   differs from the NFW collisionless halo,
   in which $\rho_\mathrm{d}\sim r^{-3}$ at large radii
   \citep{nfw1996}.
Table~\ref{table.radii} shows standard radius ratios
   for various $F$.
The radius where the circular orbit velocity is maximal
   is also given in terms of $R_\mathrm{o}/R$.
Other models' key radii were stated in
   Appendix~D of \cite{saxton2008}.

\begin{table}
\caption{%
Signature radii of perfect polytropic halos
with thermal degrees of freedom $F$
and outer truncation radius $R$.
The radii $R_1, R_2, R_3$ and $R_4$ mark
   integer steps in the density index.
The velocity of circular orbits is maximal at $R_\mathrm{o}$.
}
\begin{center}
$\begin{array}{r@{.}lr@{.}lr@{.}lr@{.}lr@{.}lr@{.}l}
\hline
\multicolumn{2}{c}{F}
&\multicolumn{2}{c}{R_1/R}
&\multicolumn{2}{c}{R_2/R}
&\multicolumn{2}{c}{R_3/R}
&\multicolumn{2}{c}{R_4/R}
&\multicolumn{2}{c}{R_\mathrm{o}/R}
\\
\hline
2&0	&0&500		&0&646		&0&729		&0&781		&0&873		\\
3&0	&0&379		&0&520		&0&614		&0&681		&0&750		\\
4&0	&0&288		&0&413		&0&507		&0&583		&0&623		\\
5&0	&0&215		&0&319		&0&407		&0&486		&0&499		\\
6&0	&0&155		&0&237		&0&313		&0&389		&0&381		\\
7&0	&0&106		&0&165		&0&225		&0&293		&0&271		\\
8&0	&0&0636		&0&101		&0&143		&0&196		&0&169		\\
9&0	&0&0285		&0&0459		&0&0668		&0&0988		&0&0782		\\
9&5	&0&0133		&0&0216		&0&0320		&0&0495		&0&0372		\\
9&9	&0&00250	&0&00408	&0&00610		&0&00986	&0&00706	\\
\hline
\end{array}$
\end{center}
\label{table.radii}
\end{table}

\section{equilibrium properties}
\label{s.stationary}

\begin{figure*}
\begin{center}
\begin{tabular}{ccc}
 \includegraphics[width=84mm]{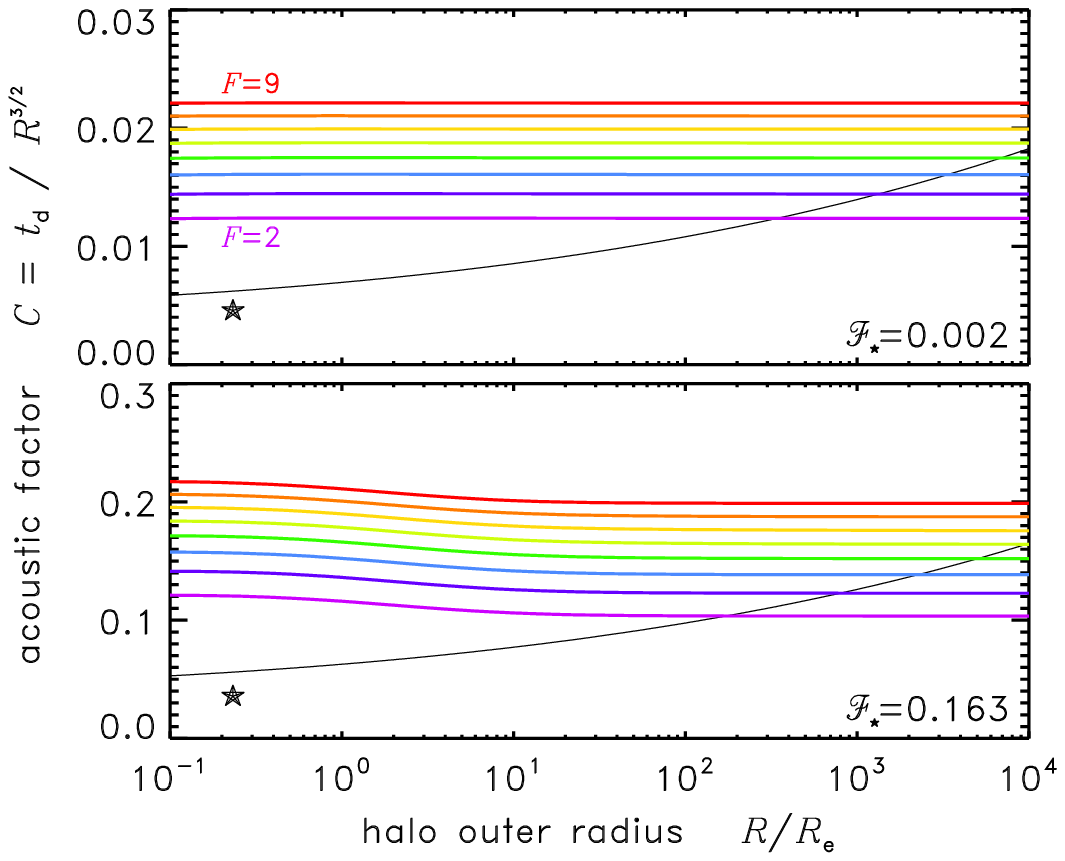}
&\includegraphics[width=84mm]{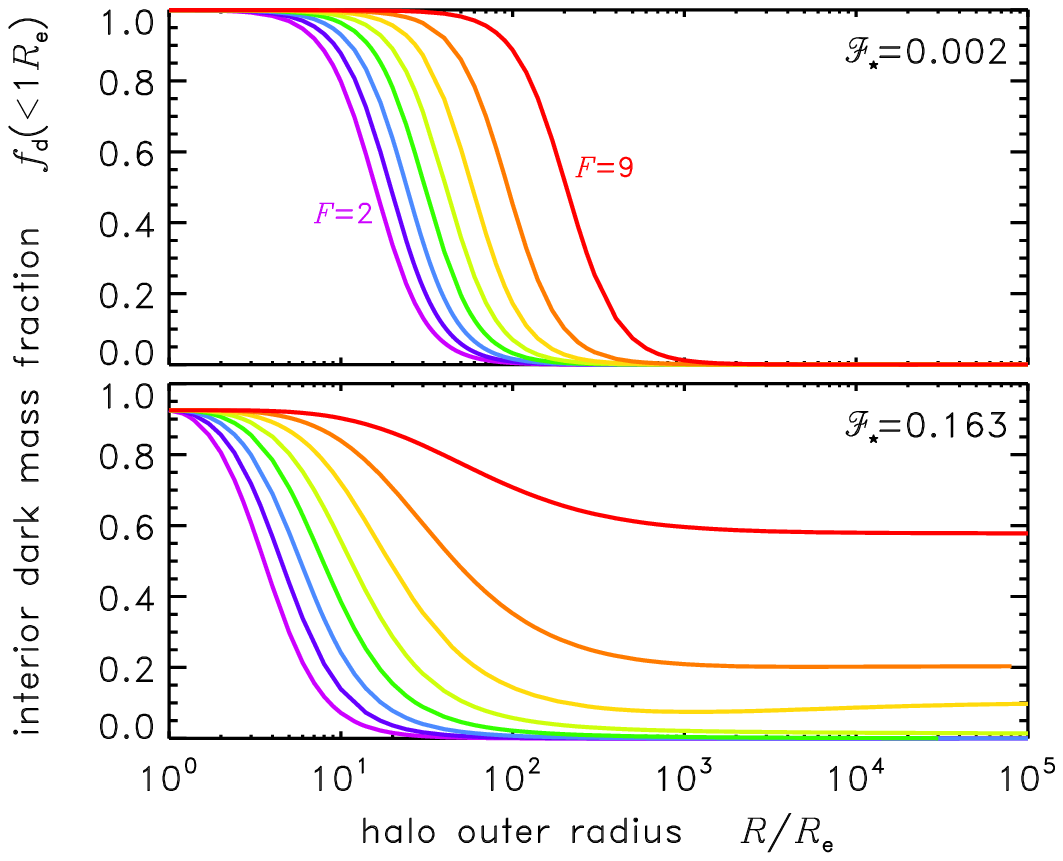}
\end{tabular}
\end{center}
\caption{%
Scaling properties of solutions
   for various halo radii ($R$, horizontal axis),
   different $F$ values (coloured curves)
   and stellar mass fractions 
   (${\mathcal F}_\bigstar=0.002$ and $ 0.163$).
The left column shows the factor $C\equiv t_{\mathrm d}/R^{3/2}$
   in the relation between acoustic crossing time
   and halo outer radius.
The black curve shows the crossing timescale
   of stars (on isotropic orbits) within the halo
   ($t_\bigstar$ replacing $t_{\mathrm d}$).
The right column shows the dark mass fraction within $1R_{\mathrm e}$
   (the half-light radius of the stars).
   for models of specified halo outer radius $R$.
}
\label{fig.steady.time}
\label{fig.conc1}
\end{figure*}

The acoustic crossing time of the dark matter halo
   (\S\ref{s.acoustic})
   scales with the halo radius as
   $t_{\mathrm d}=C\,R^{3/2}$,
   where the factor $C=C(R,F)$ varies slowly in $R$.
For a fixed stellar mass fraction (${\mathcal F}_\bigstar$)
   the factor $C$ is slightly larger for greater $F$.
The global mass ratio has a greater effect on $C$
   than $R$ or $F$ do.
The left panels of Figure~\ref{fig.steady.time}
   show the range of values obtained
   for DM-dominated models,
   and cosmic mean composition.
The crossing time associated with the stars
   is shorter than for dark matter,
   for all models except those with very large radii.
This is a result of the coldness of the dark matter
   in layers near the halo outer surface,
   where the stellar density profile is assumed to continue outwards gradually.
Within the half-light radius of the stars,
   the dark matter has a higher velocity dispersion.
Consequently, disturbances traverse the central regions
   faster via dark matter than via stars;
   and faster via stars in the outskirts.

The right panels of Figure~\ref{fig.conc1}
   show the dark mass fraction
   within $1R_{\mathrm e}$
   as a function of the outer radius of the halo.
For models with $F<6$,
   these fractions decline rapidly with $R$.
For models with $F>6$ the dark mass fraction
   within these central regions is not very sensitive
   to the outer extent of the halo;
   it is effectively determined by
   the global ratio of dark to stellar matter.
In observed systems where the dark matter fraction has been estimated
   from the kinematics of stars, planetary nebulae or globular clusters,
   the range of model $f_{\rm d}(<1R_{\rm e})$ and $f_{\rm d}(<5R_{\rm e})$
   might be used graphically to constrain the likeliest
   $(F,{\mathcal F}_\bigstar,R)$ parameters of the galaxy
   as a preliminary step to seismic modelling.

\section{Jeans equation for stellar pressure}
\label{appendix.energy}

The stars are described by a velocity distribution function
   in spherical coordinates,
   $f=f(r,\theta,\phi,v_r,v_\theta,v_\phi)$,
   where mean observables at any radius $r$
   are obtained by integration over velocity-space,
   $\langle{q}\rangle=\iiint fq\,\mathrm{d}^3v$.
(This appendix will omit the $\bigstar$ subscripts.)
In spherical coordinates, the collisionless Boltzmann equation is
\begin{eqnarray}
{{\partial f}\over{\partial t}}
+v_r{{\partial f}\over{\partial r}}
+{{v_\theta}\over{r}}{{\partial f}\over{\partial\theta}}
+{{v_\phi}\over{r\sin\theta}}{{\partial f}\over{\partial\phi}}
\nonumber\\
+\dot{v}_r{{\partial f}\over{\partial v_r}}
+\dot{v}_\theta{{\partial f}\over{\partial v_\theta}}
+\dot{v}_\phi{{\partial f}\over{\partial v_\phi}}
\hspace{-2mm}&=&\hspace{-2mm}0
\label{eq.cbe}
\end{eqnarray}
with the derivative terms
\begin{eqnarray}
	\dot{v}_r
	&\hspace{-3mm}=&\hspace{-3mm}
	{{v_\theta^2+v_\phi^2}\over{r}}
	-{{\partial\Phi}\over{\partial r}}
\\
	\dot{v}_\theta
	&\hspace{-3mm}=&\hspace{-3mm}
	{{v_\phi^2\cot\theta-v_rv_\theta}\over{r}}
	-{1\over{r}}{{\partial\Phi}\over{\partial\theta}}
\\
	\dot{v}_\phi
	&\hspace{-3mm}=&\hspace{-3mm}
	{{-v_\phi(v_r+v_\theta\cot\theta)}\over{r}}
	-{1\over{r\sin\theta}}{{\partial\Phi}\over{\partial\phi}}
\ .
\end{eqnarray}
Multiplying (\ref{eq.cbe}) by an observable quantity $q$
   and taking means over the velocity-space
   yields moment equations of the type found by \cite{jeans1915}.

Consider the special case where a spherically symmetric galaxy is non-rotating:
   $\theta$ and $\phi$ are irrelevant 
   and the 
   $v_\theta$ and $v_\phi$
   velocity components are indistinguishable.
Then we have
   $f(r,v_r,v_\theta,v_\phi)=f(r,v_r,\pm v_\theta,\pm v_\phi)$.
If we consider purely radial perturbations
   in which the distribution function
   remains even with respect to the transverse velocity components,
   then some velocity moments vanish,
   (e.g. $\langle{v_\theta}\rangle=0$,
   $\langle{v_r^2v_\theta}\rangle\approx0$)
   and
   $\langle{v_rv_\theta^2}\rangle=\langle{v_rv_\phi^2}\rangle
   \approx v\sigma_\perp^2$.   
The gravitational potential $\Phi$ is spherical,
   so $\partial\Phi/\partial\theta=\partial\Phi/\partial\phi=0$,
   and the gravitational field is radial ($g=-\partial\Phi/\partial r$).
Under these simplifications, multiplying (\ref{eq.cbe})
   by $v_r^2$ and integrating over velocity-space
   obtains a time-dependent spherical Jeans equation for the radial pressure,
   in an eulerian frame:
\begin{eqnarray}
0&\hspace{-3mm}=&\hspace{-3mm}
{{\partial}\over{\partial t}}\rho\langle{v_r^2}\rangle
+{{\partial}\over{\partial r}}\rho\langle{v_r^3}\rangle
-2\rho\langle{v_r}\rangle g
\nonumber\\&&
+{{2\rho}\over{r}}\left[{
	-\langle{v_rv_\theta^2}\rangle
	-\langle{v_rv_\phi^2}\rangle
	+\langle{v_r^3}\rangle
	+{{\langle{v_r^2v_\theta}\rangle\cot\theta}\over{2}}
}\right]
\ .
\label{eq.jeans.energy}
\end{eqnarray}
If we assume that our radial perturbations are small enough
   to preserve a zero skewness
   of the velocity distribution in respect to $v_r$,
   (e.g. a gaussian form for the stationary model)   
   then
   $\langle{v_r^2}\rangle=v^2+\sigma^2$
   and
   $\langle{v_r^3}\rangle=v(v^2+3\sigma^2)$.
This step is similar to the truncation of moment equations
   in the one-dimensional slab models of \cite{louis1992},
   but more generally allows pressure anisotropy and radial flows.
Then (\ref{eq.jeans.energy}) converts to
   the ``lagrangian frame''
   (following the local mean velocity)
   by way of
\begin{equation}
	{\mathrm{d}\over{\mathrm{d}t}}\equiv
	{\partial\over{\partial t}}
	+{\mathbf v}\cdot\nabla
	\ ,
\end{equation}
to obtain
\begin{equation}
	0={{\mathrm{d}P}\over{\mathrm{d}t}}
		+2v^3{{\partial\rho}\over{\partial r}}
		+\left({
			3P+2\rho v^2
		}\right){{\partial v}\over{\partial r}}
		+{{2\rho v}\over{r}}\left({
			v^2+2\sigma_r^2-\sigma_\perp^2		
		}\right)
		\ .
\end{equation}

\label{lastpage}
\end{document}